%% file: main.tex
\documentclass[lettersize,journal]{IEEEtran} 
\usepackage[caption=false,font=normalsize,labelfont=sf,textfont=sf]{subfig}
\usepackage{verbatim}
\usepackage[T1]{fontenc}
\usepackage{amsfonts}
\usepackage[utf8]{inputenc}
\usepackage{makecell}
\usepackage{array}
\usepackage{tabularx}
\usepackage{threeparttable}
\usepackage{soul}
\AtBeginDocument{%
  }
\usepackage{tikz}
\usepackage{setspace}
\usepackage{mathtools}
\usepackage{amsthm}
\usepackage{color}
\usepackage{booktabs}
\usepackage{graphicx}
\usepackage{xcolor}

\usepackage{comment}
\usepackage[colorlinks,linkcolor=red,anchorcolor=blue,citecolor=blue]{hyperref}
\usepackage{textcomp}
\usepackage{stfloats}
\usepackage{url}
\usepackage{cite}
\usepackage{marvosym}
\usepackage{ifsym}
\usepackage{pifont}
\usepackage{marginnote}
\usepackage{xspace}
\usepackage{stmaryrd,amsmath,amssymb}
\SetSymbolFont{stmry}{bold}{U}{stmry}{m}{n} 
\usepackage{adjustbox}
\usepackage{multirow, diagbox}
\usepackage{algorithm}
\usepackage{algorithmicx}
\usepackage{algpseudocode}
\usepackage{enumitem}
\usepackage[most]{tcolorbox}

\usetikzlibrary{positioning,arrows.meta,fit,calc,backgrounds}

\input{macros}

\begin{document}
\bstctlcite{IEEEexample:BSTcontrol}

\title{Flexibility Has a Price: Quantifying Modality Policies in Maliciously Secure Multi-Biometric Authentication}

\author{ 
Fuyi~Wang, Wang Zhang, Neo Xu, Chao Chen, Kok-Leong Ong, Jianying Zhou~\IEEEmembership{Fellow,~IEEE}
\IEEEcompsocitemizethanks{
\IEEEcompsocthanksitem This research is supported by the National Research Foundation, Singapore and Infocomm Media Development Authority under its Trust Tech Funding Initiative (DTC-T2FI-CFP-0002). Any opinions, findings and conclusions or recommendations expressed in this material are those of the author(s) and do not reflect the views of National Research Foundation, Singapore and Infocomm Media Development Authority.
\IEEEcompsocthanksitem F. Wang, N. Xu, C. Chen, K.L. Ong are with School of Accounting, Information Systems and Supply Chain, RMIT University, VIC, Australia. E-mail: \{fuyi.wang,neo.xu,chao.chen,kok-leong.ong2\}@rmit.edu.au.
\IEEEcompsocthanksitem W. Zhang is with the School of Artificial Intelligence, Beijing Normal University, Beijing, China. E-mail: zhangwang@mail.bnu.edu.cn.
\IEEEcompsocthanksitem J. Zhou is with the Centre for Research in Cyber Security, Singapore University of Technology and Design, Singapore. E-mail: jianying\_zhou@sutd.edu.sg. 
}}

\IEEEtitleabstractindextext{%
\begin{abstract}
\input{sections/00-abstract}
\end{abstract}
}

\maketitle
\IEEEdisplaynontitleabstractindextext
\IEEEpeerreviewmaketitle

\input{sections/01-introduction}

\input{sections/02-related-work}

\input{sections/03-preliminaries}

\input{sections/04-model}

\input{sections/05-flexibility-cost}

\input{sections/06-building-blocks}

\input{sections/07-design}

\input{sections/08-analysis}

\input{sections/09-evaluation}

\input{sections/11-conclusion}



\bibliographystyle{IEEEtran}
\bibliography{refs}

\end{document}

%% file: macros.tex
\newcommand{\sysname}{\textsc{SuperMBA}\xspace}

\newcommand*{\circled}[1]{\lower.7ex\hbox{\tikz\draw (0pt, 0pt)%
    circle (.4em) node {\makebox[0.4em][c]{\small #1}};}}
\newcommand*\emptycirc[1][1ex]{\tikz\draw[thick] (0,0) circle (3pt);}
\newcommand*\halfcircc[1][1ex]{\tikz{
    \draw[thick] (0,0) circle (4pt);
    \fill[black] (0,0) -- ++(0, 4pt) arc[start angle=90, end angle=-90, radius=4pt] -- cycle;
}}
\newcommand*\halfcirc[1][1ex]{\tikz{
    \draw[thick] (0,0) circle (5pt);
    \fill[black] (0,0) -- ++(0, 5pt) arc[start angle=90, end angle=-90, radius=5pt] -- cycle;
}}
\newcommand*\fullcirc[1][1ex]{\tikz\filldraw[fill=black] (0,0) circle (3pt);}

\newtcolorbox{mybox}[2][]{%
  enhanced,
  title        = {#2},
  attach boxed title to top left={xshift=+3mm,yshift*=-3mm},
  colback      = white,
  colframe     = black,
  fonttitle    = \bfseries,
  fontupper    = \small,
  fontlower    = \small,
  colbacktitle = black!3!white,
  coltitle     = black,
  #1
}
\newtcbox{\xmybox}[1][red]{on line, arc=7pt,colback=#1!10!white,colframe=#1!50!black, before upper={\rule[-3pt]{0pt}{10pt}},boxrule=1pt, boxsep=0pt,left=6pt,right=6pt,top=1pt,bottom=1pt}

\colorlet{pink}{red!20}
\definecolor{blue0}{HTML}{C0D3E3}
\definecolor{blue1}{HTML}{9299C1}
\definecolor{blue2}{HTML}{9170AB}

\floatname{algorithm}{Algorithm}

\theoremstyle{plain}
\newtheorem{theorem}{Theorem}
\newtheorem{lemma}{Lemma}
\newtheorem{proposition}{Proposition}
\newtheorem{corollary}{Corollary}
\newtheorem{definition}{Definition}
\theoremstyle{remark}
\newtheorem{remark}{Remark}

%% file: sections/00-abstract.tex
Multi-biometric authentication improves robustness and accessibility by combining information from multiple biometric modalities. Existing research has primarily focused on secure multimodal authentication, feature fusion, and recognition performance, while paying comparatively little attention to the security implications of modality-flexible authentication policies. At the same time, most systems provide limited cryptographic protection for biometric data during authentication.

This paper addresses both challenges. We first study modality flexibility as a security decision rather than merely a usability feature. Each accepted modality subset creates an authentication entry point, and the resulting risk depends on the joint behavior of all accepted subsets and the attempt budget. We introduce a framework for quantifying this security cost and develop a policy-selection procedure that identifies policies satisfying a target risk level while supporting users who can present only a subset of modalities.
We then present \sysname, a two-server privacy-preserving authentication system that enforces the selected policy without revealing biometric templates or the modality subset used in an authentication attempt. The protocol provides verifiable policy enforcement even when one server is malicious and executes the same computation irrespective of the chosen policy or modality subset.
Experiments on the three-modality MULB+ dataset show that allowing all modality subsets increases impostor risk from 1.21\% for the weakest entry to 4.42\% for the union of entries. On VidTIMIT, the increase is substantially smaller and is not statistically resolved with the available identities. \sysname completes an authentication attempt in only {6.42} ms on MULB+ and {6.03} ms on VidTIMIT.

%% file: sections/01-introduction.tex
\section{Introduction}
\label{sec:intro}

\IEEEPARstart{B}{}iometric authentication has become a fundamental component of modern digital identity systems, offering a convenient alternative to passwords and tokens while establishing a direct link between a user and their credentials~\cite{wang2026generating,wang2025SEBioID}. Unlike passwords, however, biometric traits cannot be revoked once compromised. Protecting biometric data during enrollment, storage, and authentication has therefore become a long-standing research challenge. A substantial body of work has shown how authentication can be performed without revealing biometric templates, using techniques such as homomorphic encryption, garbled circuits, and secure multi-party computation~\cite{pradel2021privacy,wang2025privacy,blanton2024privacy,tan2023mpcauth,wang2022towards}. Some studies further provide security against malicious adversaries that arbitrarily deviate from prescribed protocols~\cite{bassit2021fast,cheng2024nomadic,wang2025flame}. At the same time, the use of multiple biometric modalities has received increasing attention. By combining information from different traits, multi-biometric systems can achieve higher robustness than unimodal systems~\cite{zhao2024deep}. They can also improve accessibility and inclusiveness, since some users may be temporarily unable to present one because of sensor failure, injury, illness, or environmental conditions~\cite{ross2003information,wang2025flame}. Consequently, many deployed or proposed systems allow users to authenticate with whichever subset of modalities they can present.

Existing research has primarily focused on secure multimodal authentication, feature fusion, and recognition performance~\cite{zhao2024deep,talreja2020deep}. Comparatively little attention has been paid to the authentication policy itself, namely the question of which modality combinations should be accepted. In most work, modality flexibility is treated purely as a usability feature. Yet allowing additional modality combinations is fundamentally a security decision. Every accepted subset constitutes an authentication entry point, and an attacker does not need to defeat the strongest combination of modalities; it only needs one entry that admits it. The resulting risk depends not only on the performance of individual subsets but also on their joint behavior and on the number of attempts available to the attacker. These quantities cannot be inferred from the per-subset error rates that multi-biometric systems usually report. 

Additionally, protecting biometric data introduces a second challenge. Once biometric templates and authentication inputs are hidden within a secure computation, the servers executing the protocol can no longer observe which modality subset a user presents. Consequently, a policy over modality subsets must itself be enforced inside the secure computation. Revealing the subset to the servers would disclose information about which modalities a user can or cannot provide, potentially exposing sensitive information about temporary physical conditions or disabilities. Existing privacy-preserving biometric authentication systems largely focus on protecting templates and matching scores, but do not provide a systematic treatment of policy enforcement under modality flexibility.

This paper asks:
\begin{quote}
\emph{How does modality flexibility affect authentication security, and how can flexible authentication policies be enforced while preserving the privacy of biometric data and modality availability?}
\end{quote}


To answer these questions, we first analyze the security implications of modality flexibility and develop a framework for quantifying the risk introduced by allowing multiple modality combinations. Building on this analysis, we present a policy-selection procedure that translates security and usability requirements into concrete authentication policies. We then design \sysname, which incorporates the policy into a functionality jointly computed by two non-colluding servers, at most one of them malicious, so that authentication policies are enforced without revealing the presented subset and authentication decisions are verified before release. Finally, we empirically evaluate both the risk analysis and the proposed system design.
Our contributions are the following.
\begin{itemize}[leftmargin=*,topsep=0pt]
\item We present \sysname, a privacy-preserving multi-biometric authentication system that supports flexible authentication policies while explicitly accounting for their security implications. We formalize authentication risk under modality flexibility and provide a policy selection procedure that identifies policies satisfying target security and usability requirements. 
\item We design efficient two-party authentication protocols over authenticated secret sharing and function secret sharing. Our protocols hide modality availability from the servers, support policy enforcement within secure computation, and remain secure against one malicious server. 
\item We prove that \sysname securely realizes the desired authentication functionality and provides modality-subset privacy. Experiments on MULB+ and VidTIMIT demonstrate that \sysname faithfully reproduces plaintext authentication decisions while achieving low latency and practical deployment costs.
\end{itemize}

\paragraph*{Relation to the conference version} This paper extends our ACSAC paper~\cite{wang2025flame}, which 
presented maliciously secure two-party protocols for privacy-preserving biometric matching. We retain its two-server architecture with a trusted dealer, as well as its sum-then-open inner-product and authenticated comparison protocols, while introducing moderate communication/computation optimizations. 
Also, 
this paper addresses the broader problem of modality-flexible biometric authentication. Specifically, we introduce a security framework for analyzing the risk of modality flexibility, a policy-selection procedure that balances security and usability requirements, a policy-gated authentication functionality, and a protocol that privately enforces flexible authentication policies without revealing the presented modality subset. We further provide a formal security analysis, an itemized cost analysis, and a comprehensive experimental evaluation. 
For these reasons, the performance, accuracy, and security analyses reported in this paper supersede those of the conference version.

%% file: sections/02-related-work.tex
\section{Related Work}
\label{sec:related}

The question of Section~\ref{sec:intro} sits between two lines of work: maliciously secure privacy-preserving biometric authentication, which protects biometric features and the matching process; and the security of multi-biometric fusion, which studies how an attacker may exploit multiple modalities.  

\paragraph*{Privacy-preserving biometric authentication}
Early schemes evaluate the matching function under homomorphic encryption at a single server; with fully homomorphic encryption the cost of a naive worst-case evaluation is prohibitive~\cite{pradel2021privacy}. Two non-colluding servers running a mixed secret-sharing and garbled-circuit protocol reduce the cost but are assumed semi-honest~\cite{treiber2019privacy}. In the client--server setting, zero-knowledge proofs let either party be malicious~\cite{bassit2021fast}, and a helper server and garbled circuits allow the client to be malicious throughout while it stores no secrets~\cite{blanton2024privacy}. In the outsourced setting, Nomadic combines secret sharing with function secret sharing and SPDZ$_{2^k}$-style MACs so that one of two servers may be malicious, and normalizes the inputs inside the protocol so that a malicious client cannot submit an ill-formed vector~\cite{cheng2024nomadic}; FLAME follows the same two-server model~\cite{wang2025flame} and improves the efficiency of secure computations. 
Each of these systems computes a decision under a matching rule fixed in advance; which modalities a user may present, and how many attempts it may make, are decided outside the functionality they protect.

\paragraph*{Multi-biometric fusion and the choice of modalities}
Modalities can be fused at the feature, score, or decision level~\cite{ross2003information}; score-level fusion requires normalizing heterogeneous scores, and subsets of modalities have even been proposed as a liveness challenge chosen by the system~\cite{zhao2024deep}. Protection of multi-modal templates has been studied in its own right, for example by deep hashing of fused face and iris features~\cite{talreja2020deep}
. 
Some template-protection schemes fix the combination of characteristics in the design: Zhao et al.\ require both face and iris for every authentication~\cite{zhao2024deep}, and Osorio-Roig et al.\ evaluate fixed combinations of two or three characteristics~\cite{osorio2024privacy}. Privacy-preserving systems that handle several biometric sources reveal little beyond a decision: Janus deduplicates aid recipients against several sources while revealing a single bit~\cite{edalatnejad2024janus}, and MPCAuth hides users' authentication profiles, including biometric features, across servers~\cite{tan2023mpcauth}. Neither paper describes the set of modalities a user presents as an input to a policy.


That set is an attack surface, and the work closest to ours analyzes exactly that. Murakami et al.\ observed that likelihood-ratio score fusion can serve users who cannot present some modality by setting the missing log-likelihood ratios to zero, defined the \emph{modality selection attack} in which an impostor presents only the modalities whose log-likelihood ratios are positive, and showed that accuracy degrades even when genuine users adopt the same strategy~\cite{murakami2016info}
. These analyses use an aggregate information-theoretic measure, their remedy acts on templates, and they do not choose allowed subsets, thresholds, and an attempt budget against a risk target while meeting usability requirements per condition.

\begin{table*}[!t]
\centering
\caption{Systems closest to \sysname. ``Mal.'' = malicious; n/s = not stated in the paper. HE = homomorphic encryption (FHE = fully, SHE = somewhat); ZKP = zero-knowledge proofs; SS = secret sharing; FSS = function secret sharing; MPC = secure multi-party computation; MAC = message authentication code; TEE = trusted execution environment.
}
\label{tab:related}
\small
\resizebox{\textwidth}{!}{%
\begin{tabular}{@{}llllllll@{}}
\toprule
System & Parties & Corruption & Privacy technique & Task & Modalities & Subset policy & Subset hidden \\
\midrule
Pradel \& Mitchell~\cite{pradel2021privacy} & client, server & semi-honest server & FHE (TFHE) & $1{:}1$ & face & -- & -- \\
Treiber et al.~\cite{treiber2019privacy} & client, 2 servers & semi-honest servers & MPC & $1{:}1$ & voice & -- & -- \\
Bassit et al.~\cite{bassit2021fast} & client, server & mal.\ client or server & HE + ZKP & $1{:}1$ & signature, face & -- & -- \\
Blanton \& Murphy~\cite{blanton2024privacy} & client, 2 servers & mal.\ client (+ helper) & MPC with helper server & $1{:}1$ & fingerprint & -- & -- \\
Nomadic~\cite{cheng2024nomadic} & client, dealer, 2 servers & mal.\ server & SS + FSS + MAC & $1{:}1$ & voice & -- & -- \\
FLAME~\cite{wang2025flame} & client, dealer, 2 servers & mal.\ servers & SS + FSS + MAC & $1{:}N$ & face & -- & -- \\
Talreja et al.~\cite{talreja2020deep} & client, server & no formal model & deep hashing + secure sketch & $1{:}1$ & face+iris & n/s & n/s \\
Zhao et al.~\cite{zhao2024deep} & client, server & semi-honest server & deep hashing   & $1{:}1$ & face+iris & n/s & n/s \\
Janus~\cite{edalatnejad2024janus} & 2 parties & semi-honest & MPC; SHE + MPC; TEE & dedup.\ ($1{:}N$) & finger, iris, face & n/s & n/s \\
MPCAuth~\cite{tan2023mpcauth} & client, $N$ servers & $\le N{-}1$ mal.\ servers & MPC & $1{:}1$ & face (+ non-biometric) & n/s & n/s \\
\midrule
\sysname & client, dealer, 2 servers & $\le1$ mal.\ server & SS + FSS + MAC & $1{:}1$ & face, iris, hand; face, voice & yes & yes \\
\bottomrule
\end{tabular}}
\end{table*}

Table~\ref{tab:related} makes the gap concrete: the maliciously secure systems are single-modality and have no subset policy to enforce, and the papers on multi-modal systems do not describe the presented subset as private or as an input to a policy.

%% file: sections/03-preliminaries.tex
\section{Preliminaries}\label{sec:prelim}

\subsection{Multi-Biometric Verification and Policies}
\label{ssec:prelim-fusion}

Let $\mathcal M$ be the set of modalities a system supports and $\mathcal S=2^{\mathcal M}\setminus\{\emptyset\}$ the family of non-empty subsets of modalities, so that $K=|\mathcal S|$; with three modalities, $K=7$. Each modality $m$ has a fixed feature extractor that outputs a unit vector $v_m\in\mathbb R^{d_m}$, and fusion models are required for multi-biometric subsets. For simplicity, we denote by $\phi_S(x_S)$ the model executed on the client for the captured subset of modalities $S$. This notation abstracts both unimodal and multimodal cases. In particular, when multiple modalities are available, $\phi_S$ includes the corresponding feature extractors as well as the fusion model that combines their representations into a single embedding.
To prevent the modality used during authentication from being inferred from the feature dimension, all modality-specific embeddings are mapped to a common representation size. Let $ d=\max_{m\in\mathcal M} d_m $ be the largest modality dimension. Feature vectors whose original dimension is smaller than $d$ are zero-padded to length $d$. 
This prevents an observer from inferring the submitted modality subset solely from dimensionality information.
For an account $u$ presenting subset $S$ with samples $x_S$, the verification score is the inner product $s_S=\langle\phi_S(x_S),T_{u,S}\rangle$ between the fused embedding and the template $T_{u,S}$ of account $u$ for subset $S$; it is a cosine similarity because both arguments are unit norm~\cite{wang2025flame}. The attempt is accepted when $s_S\ge\tau_S$.

\begin{definition}[Flexible authentication policy]
\label{def:policy}
A policy is a triple $P=(\mathcal A,\boldsymbol\tau,B)$: an allowed family $\mathcal A\subseteq\mathcal S$, a threshold $\tau_S$ for each $S\in\mathcal A$, and a budget $B\ge1$ of admitted attempts per account and time window. An admitted attempt that presents subset $S$ with samples $x_S$ for account $u$ is accepted iff $S\in\mathcal A$ and $s_S\ge\tau_S$.
\end{definition}

\subsection{Authenticated Secret Sharing}
\label{ssec:prelim-ss}

Two servers $\mathcal P_0,\mathcal P_1$ hold secret-shared values over the ring $\mathbb Z_{2^\ell}$. As in SPDZ$_{2^k}$~\cite{cramer2018spd}, values and message authentication codes (MACs) live in the larger ring $\mathbb Z_{2^{\ell+\sigma}}$: the MAC key $\alpha\in\mathbb Z_{2^\sigma}$ is additively shared between the servers, and the MAC of a value $x$ is $\alpha x \bmod 2^{\ell+\sigma}$.

A value $x$ is held in masked form $(\Delta_x,\langle\lambda_x\rangle)$~\cite{ben2019turbospeedz}, where $\Delta_x=x+\lambda_x$ is known to both servers and the mask $\lambda_x$ is additively shared between them. The mask is input-independent and fixed during \emph{preprocessing}, which a dealer carries out without knowing the inputs to the operations it prepares. We model it as the functionality $\mathcal F_{\mathsf{Prep}}$: it samples the MAC key shares, the masks, the Beaver triples together with their offsets, and the FSS keys of Section~\ref{ssec:prelim-fss}, gives each server its part of this material, and gives the client the masks of its own inputs; no step depends on biometric data or on any future authentication attempt. The authenticated masked share $[\![x]\!]$ is the pair of masked shares of $x$ and of $\alpha x$~\cite{wang2025flame}. Linear operations are local: $[\![x]\!]+[\![y]\!]$ adds both components, and multiplication by a public constant $c$ maps $(\Delta_x,\langle\lambda_x\rangle)$ to $(c\Delta_x,c\langle\lambda_x\rangle)$; the mask must be scaled together with the value. Adding a public constant $c$ is also local: the value component becomes $(\Delta_x+c,\langle\lambda_x\rangle)$ and the MAC component $(\Delta_{\alpha x},\langle\lambda_{\alpha x}\rangle-c\langle\alpha\rangle)$, since $\alpha(x+c)=\alpha x+\alpha c$.

Multiplication consumes a Beaver triple $(A,B,C=AB)$ from preprocessing together with public offsets computed by the dealer, and opens one masked value. Opened values are verified in a batch with the MAC check of SPDZ$_{2^k}$~\cite{cramer2018spd}; Section~\ref{sec:analysis} states its error probability.

\subsection{Function Secret Sharing}
\label{ssec:prelim-fss}

Function Secret Sharing (FSS)~\cite{boyle2016fssimproved,boyle2021function} splits a function $f:\mathbb F_{\mathrm{in}}\rightarrow\mathbb F_{\mathrm{out}}$ into two share functions $(f_0,f_1)$ with $f(x)=f_0(x)+f_1(x)$ for every $x\in\mathbb F_{\mathrm{in}}$, such that each share alone reveals nothing about $f$. A common instance is the Distributed Comparison Function (DCF): for a threshold $\gamma$ and payload $\beta$, $f^{<}_{\gamma,\beta}(x)=\beta$ if $x<\gamma$ and $0$ otherwise, which lets two parties evaluate a comparison without learning $\gamma$ or the outcome individually. A standard two-party FSS construction provides:
\begin{itemize}[leftmargin=*,topsep=0pt]
  \item $\mathsf{KeyGen}(1^\lambda,f)\rightarrow(k_0,k_1)$: given a security parameter $\lambda$ and a function $f$, generates a pair of correlated keys, each defining one share function held by a different party.
  \item $\mathsf{Eval}(j,k_j,x)\rightarrow y_j$: given a party index $j\in\{0,1\}$, key $k_j$ and input $x\in\mathbb F_{\mathrm{in}}$, outputs the share $y_j=f_j(x)\in\mathbb F_{\mathrm{out}}$.
\end{itemize} 
The output is recovered as $f(x)=y_0+y_1$. Since evaluation is local and uses pre-distributed keys, FSS greatly reduces online communication.


%% file: sections/04-model.tex
\section{System Overview}
\label{sec:model}


\subsection{System Model}
\label{ssec:model-system}
\begin{figure*}[!t]
\centering
\includegraphics[width=\linewidth]{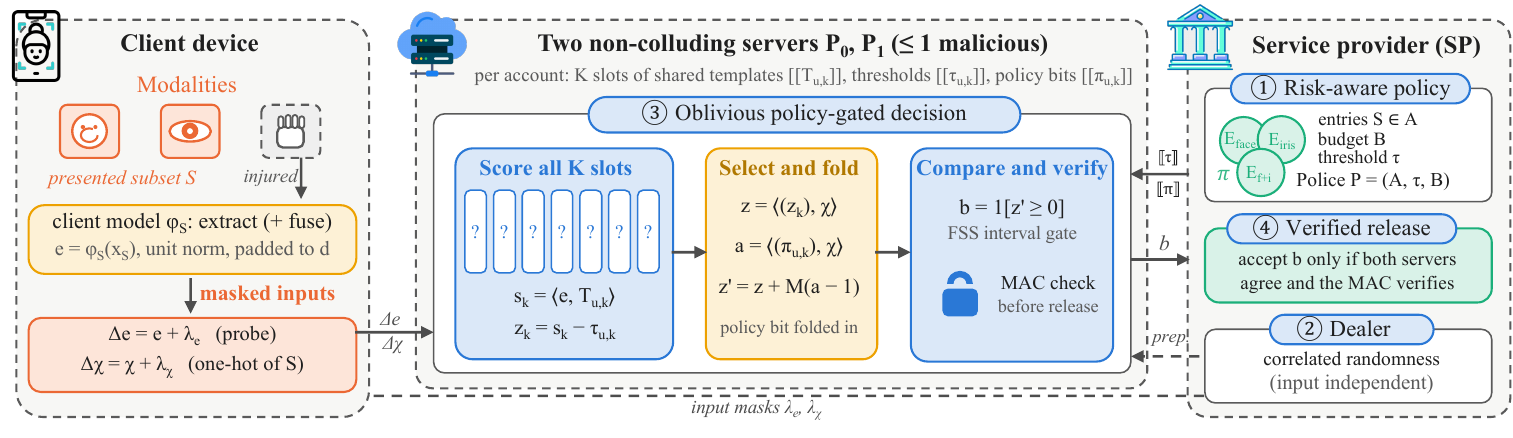}
\vspace{-6mm}
\caption{System model of \sysname. 
}
\label{fig:system}
\end{figure*}

We consider a scenario where a client attempts to log in to government digital services via Singpass using a multi-biometric authentication system, $\sysname$.
As depicted in Figure~\ref{fig:system}, $\sysname$ consists of three entities: the client $\mathcal{C}$, the service provider $\mathcal{S}$, and two servers $\mathcal{P}_0$ and $\mathcal{P}_1$.
\begin{itemize}[leftmargin=*,topsep=0pt]
\item The \emph{client device} $\mathcal C$ captures the available modalities and computes an embedding $e=\phi_S(x_S)$ for the captured subset $S$, where $\phi_S$ denotes the client-side model for $S$ (including multimodal fusion when $|S|>1$). The embedding is normalized and encoded in fixed point. With $f$ fractional bits, each coordinate is represented as $\hat e_i=\lfloor 2^f e_i\rceil$ in two's complement over $\mathbb Z_{2^\ell}$; templates are encoded identically. Scores are computed as integer inner products at scale $2^{2f}$, and thresholds use the same scale. No truncation is required, since scores are compared against thresholds and multiplied only by indicator bits.
\item The \emph{service provider} $\mathcal{SP}$ is the relying party. It assigns each account its policy (Definition~\ref{def:policy}), acts as the dealer that generates input-independent correlated randomness, admits authentication attempts, and receives the decision.
\item Two \emph{computing servers} $\mathcal P_0,\mathcal P_1$, operated by different organizations, store each account's templates, thresholds, and policy bits in authenticated secret-shared form and run the authentication protocol.
\end{itemize}
All links are private and authenticated: from $\mathcal{SP}$ to the client (input masks) and to each server (preprocessing), from the client to each server (masked inputs), and from each server to $\mathcal{SP}$ (the masked decision). $\mathcal{SP}$, which as the dealer knows the masks, must not observe what the client sends to the servers, and a server must not observe the masks that $\mathcal{SP}$ sends to the client; either would recover the embedding.

\sysname has three phases. In \emph{enrollment}, the device computes a template for every subset the user enrolls and fills the remaining slots with dummy templates, and $\mathcal{SP}$ assigns the account's policy; both are secret-shared to the servers. In \emph{preprocessing}, $\mathcal{SP}$ generates correlated randomness for future attempts, independently of any biometric data. In \emph{authentication}, the device submits a fresh embedding for a claimed identity $u$, which is public, and the servers compute the decision for $\mathcal{SP}$. The primary task is $1{:}1$ verification; Section~\ref{ssec:design-1n} discusses $1{:}N$ identification as a cost extension.

\subsection{The Policy-Gated Authentication Functionality}
\label{ssec:model-func}

Index the $K$ non-empty subsets of $\mathcal M$ by $k\in[K]$. For every account $u$ the functionality stores the state
\[
\mathsf{st}_u=\big(T_{u,k},\ \hat\tau_{u,k},\ \pi_{u,k}\big)_{k\in[K]},
\]
where $\pi_{u,k}=1$ iff subset $k$ belongs to the allowed family $\mathcal A_u$ of the account's policy. A subset that the user has not enrolled holds a dummy template---a random unit vector---with $\pi_{u,k}=0$, so every account has exactly $K$ slots of the same form whatever its enrollment and policy. The templates $T_{u,k}$ come from the client and the thresholds $\hat\tau_{u,k}$ and policy bits $\pi_{u,k}$ from $\mathcal{SP}$; neither learns the other's inputs. The functionality is defined for unit-norm encodings and thresholds in $[-1,1]$, the inputs an honest client application and a correctly configured $\mathcal{SP}$ produce; a subset that a policy should never accept is excluded by its policy bit, not by an out-of-range threshold. Figure~\ref{fig:func} gives the functionality.

\begin{figure}[!t]
\small
\begin{mybox}{Functionality $\mathcal F_{\mathsf{Auth}}$}
\textbf{Parameters:} slots $K$, dimension $d$, budget $B$ per account, fixed-point encoding of Section~\ref{ssec:model-system}.\\[2pt]
\textbf{Enroll}$(u)$: receive encoded unit-norm templates $(T_{u,k})_{k\in[K]}$ from $\mathcal C$ and $(\tau_{u,k},\pi_{u,k})_{k\in[K]}$ with $\tau_{u,k}\in[-1,1]$ from $\mathcal{SP}$; store $\mathsf{st}_u$ with encoded thresholds $\hat\tau_{u,k}=\lfloor2^{2f}\tau_{u,k}\rceil$; send $u$ to $\mathcal P_0,\mathcal P_1$. \\[2pt]
\textbf{Update}$(u)$ from $\mathcal{SP}$ replaces $(\tau_{u,k},\pi_{u,k})_k$, re-encodes the thresholds, increments the policy version $v_u$, and sends $(u,v_u)$ to $\mathcal P_0,\mathcal P_1$. At the start of each authentication, $a_u\gets0$ for all $u$.\\[2pt]
\textbf{Authenticate}$(u)$:
\begin{enumerate}[leftmargin=*,topsep=0pt]
\item \emph{Admission.} If $a_u\ge B$, send \textsf{locked} to $\mathcal{SP}$ and stop. Otherwise set $a_u\gets a_u+1$, pick a fresh $\mathsf{sid}$, and send the admission metadata $(u,v_u,\mathsf{sid})$ to $\mathcal P_0,\mathcal P_1$.
\item \emph{Input.} Receive $(e,k)$ from $\mathcal C$, with $e$ an encoded unit-norm embedding and $k\in[K]$.
\item \emph{Decision.} $b\gets\pi_{u,k}\land\mathbf 1\big[\langle e,T_{u,k}\rangle\ge\hat\tau_{u,k}\big]$, an inner product of encodings compared with an encoded threshold.
\item \emph{Output.} Send $(\mathsf{sid},b)$ to $\mathcal{SP}$.
\end{enumerate}
\textbf{Abort.} The adversary may replace the output to $\mathcal{SP}$ by $(\mathsf{sid},\bot)$, which all parties learn; the attempt remains counted.
\end{mybox}
\caption{{The functionality realized by \sysname.} 
}
\label{fig:func}
\end{figure}

Three features of $\mathcal F_{\mathsf{Auth}}$ go beyond a thresholded similarity check, and each answers an attack on the policy rather than on the template. First, the subset is an \emph{input} of the functionality rather than a parameter of the protocol: if the servers ran a different computation for each subset, the transcript would reveal which modalities a user could present, which can reflect an injury or a medical condition. Second, the budget lives inside the functionality and is charged at admission. Because the risk of a flexible policy depends on how many attempts an impostor gets (Section~\ref{sec:cost}), an attempt must count whether it succeeds, fails, or is aborted. Third, the decision is the only output, and it goes only to $\mathcal{SP}$.

\subsection{Threat Model}
\label{ssec:model-threat}

Secure computation protects templates, inputs, and the integrity of the decision against the servers; it does nothing against an impostor who presents the wrong person to an honestly operated system.

\paragraph*{Computation adversary}
A probabilistic polynomial-time adversary statically corrupts at most one of $\mathcal P_0,\mathcal P_1$ and may deviate arbitrarily from the protocol. The servers do not collude. $\mathcal{SP}$ generates correlated randomness correctly and does not collude with either server, an assumption shared with existing studies~\cite{wang2025flame,cheng2024nomadic}. The client application runs the protocol honestly. We require security with abort: a corrupted server may make an attempt fail, but it cannot make $\mathcal{SP}$ accept a wrong decision, and it learns nothing beyond the admission metadata, policy updates, timing, and aborts. 

\paragraph*{Authentication adversary}
An impostor targets the account of a user $u$. It controls what is presented to an honest client application---whose biometric samples, and which subset of modalities---and it observes the decision of each attempt it is admitted to make in a window. It has no access to templates, scores, or server state. We distinguish impostors by what they know when they choose a subset and by whether they may reuse or recapture samples:
\begin{itemize}[leftmargin=*,topsep=0pt]
\item a \emph{fixed-entry} impostor $\mathsf A_{\mathrm{fix}}(S)$ always presents subset $S$;
\item an \emph{enumerator} $\mathsf A_{\mathrm{enum}}$ keeps one bundle of samples and changes the subset after each rejection, stopping at the first acceptance;
\item a \emph{resampler} recaptures fresh samples between attempts, with a fixed or changing subset.
\end{itemize}

\paragraph*{Out of scope}
Presentation attacks and liveness detection; impostors that synthesize samples close to many templates or adapt their inputs to the decision bit; integrity of enrollment; a compromised client application, which could in particular submit a stolen embedding; confidentiality of the feature extractors and fusion models; 
and denial of service, including an adversary that exhausts a victim's budget. For the distances studied by Pagnin et al., an adversary that can submit chosen inputs, knows the threshold, and holds one accepted input can recover the template from accept/reject decisions alone~\cite{pagnin2014leakage}; the budget limits the number of such queries, but we do not quantify the risk of adaptive or synthetic-input attacks.

\subsection{Design Goals}
\label{ssec:model-goals}

\sysname has four design goals. \textbf{G1}~(policy enforcement): $\mathcal{SP}$ obtains the decision of $\mathcal F_{\mathsf{Auth}}$ on the encoded inputs, including the policy bit and the budget, or $\bot$ (Theorem~\ref{thm:main}). \textbf{G2}~(subset privacy): the view of a corrupted server can be simulated without $k$, and the only information about $k$ available to $\mathcal{SP}$ is $b$ (Corollary~\ref{cor:subset}); Remark~\ref{rem:tv} shows that nothing stronger is possible. \textbf{G3}~(verifiability): a deviating server makes $\mathcal{SP}$ accept a wrong decision only with probability negligible in the statistical parameter $\sigma$ and the computational parameter $\kappa$ (Theorem~\ref{thm:main}). \textbf{G4}~(predictable cost): the server-to-server online communication of an attempt is independent of the dimension $d$ and linear in $K$; computation, input traffic, and preprocessing are reported separately.

%% file: sections/05-flexibility-cost.tex
\section{The Security Cost of Modality Flexibility}
\label{sec:cost}

The functionality of Section~\ref{sec:model} enforces whatever policy it is given. This section is about choosing that policy. It defines the quantities that describe the risk of a flexible policy, states what can be said about them in general, and turns a risk target and a set of usability requirements into a concrete policy ---or into the finding that the searched policies cannot meet them. 

\subsection{Entry Points and Their Joint Behavior}
\label{ssec:cost-risk}

Under a policy $P=(\mathcal A,\boldsymbol\tau,B)$ (Definition~\ref{def:policy}), every allowed subset $S\in\mathcal A$ is an \emph{entry point}. Consider one impostor attempt against an account. Let $E_S$ be the event that entry $S$ accepts the impostor's samples, and $p_S=\Pr[E_S]$; for the zero-effort impostor, who presents genuine samples of a different person, $p_S=\mathrm{FMR}_S(\tau_S)$. Per-subset evaluation reports the vector $(p_S)_{S\in\mathcal A}$, or its EER counterpart. The deployed system is exposed to something else: whether the impostor gets in through \emph{some} entry within its budget. We use four quantities, all defined on a common distribution of impostor trials in which every subset is scored on the same samples:
\begin{gather*}
R_{\mathrm{weak}}=\max_{S\in\mathcal A}p_S,\qquad
U=\Pr\Big[\bigcup_{S\in\mathcal A}E_S\Big],\\
\Delta_{\mathrm{sel}}=U-R_{\mathrm{weak}},\qquad
\Sigma=\sum_{S\in\mathcal A}p_S-U.
\end{gather*}
$U$ means union risk, $\Delta_{\mathrm{sel}}$ measures how much more than the weakest entry the entries let in together; $\Sigma$ measures how much they overlap. For an impostor class $\mathsf A$ from Section~\ref{ssec:model-threat} making at most $Q\le B$ attempts in a window, the risk of $P$ is $\rho_Q(P;\mathsf A)=\Pr[\exists\,t\le Q:\ b_t=1]$.

\begin{proposition}[Risk of a flexible policy]
\label{prop:ordering}
On a common trial distribution,
\[
R_{\mathrm{weak}}\ \le\ U\ \le\ \sum_{S\in\mathcal A}p_S.
\]
Every impostor that reuses one sample bundle has $\rho_Q\le U$ for all $Q$. An enumerator whose order of entries is fixed independently of the trial attains $\Pr[\bigcup_{S\in\mathcal Q}E_S]$, where $\mathcal Q$ is the set of its first $Q$ entries; with $Q\ge|\mathcal A|$ it attains $U$.
\end{proposition}

The first inequality holds because every $E_S$ is contained in the union; the second is the union bound. A bundle-reusing impostor succeeds only on a trial in which some entry accepts, hence $\rho_Q\le U$; for a fixed order, a rejection reveals only that the tried entry failed, so success within $Q$ attempts is the union over the first $Q$ entries, and an order that adapts to observable information can do better at small $Q$ but never beyond $U$. These relations are elementary, and we do not claim them as results. Their role is to separate an impostor who replays one bundle across subsets, bounded by and with enough attempts reaching the union, from one who recaptures samples between attempts, whose risk grows with $Q$ even under a single entry and must be compared with a legitimate retry policy under the same budget. Where the risk of bundle reuse falls between $R_{\mathrm{weak}}$ and $U$ is decided by the overlap of the entries' acceptance sets, which per-subset scores summarized separately cannot reveal: if the other entries accept only impostors that the weakest one accepts, $\Delta_{\mathrm{sel}}=0$; if the acceptance sets are disjoint, $\Sigma=0$ and the cost is the sum.

The quantity we measure is the realizable gain of a choosing impostor over the entry that calibration data identify as weakest, at the same budget,
\[
D_Q=\rho_Q(P;\mathsf A)-\rho_Q\big(P;\mathsf A_{\mathrm{fix}}(\hat S)\big),\qquad \mathsf A=\mathsf A_{\mathrm{enum}}.
\]
For bundle reuse, $D_Q\le U-p_{\hat S}$, which equals $\Delta_{\mathrm{sel}}$ when $\hat S$ is the weakest entry in the population and exceeds it otherwise; if the other entries' acceptance sets are nested in that of $\hat S$, then $D_Q\le0$. Comparing $Q$ impostor attempts with one legitimate attempt would measure the budget, not the flexibility.

\subsection{Choosing a Policy}
\label{ssec:cost-select}

Besides the risk target, the choice of a policy is constrained by the users whom flexibility is meant to serve. A \emph{usability condition} $A\subseteq\mathcal M$ describes a legitimate user who can present only the modalities in $A$, such as a user with an injured hand or a masked face. Under a policy with allowed family $\mathcal A$, such a user presents $r_{\mathcal A}(A)$, the first subset of $A$ allowed by $\mathcal A$ in a fixed preference order---larger subsets first, ties broken by a fixed index. We write $\mathrm{GAR}_A(P)$ for the probability that a user of condition $A$ is accepted in a single attempt with $r_{\mathcal A}(A)$; if $\mathcal A$ allows no subset of $A$, then $\mathrm{GAR}_A(P)=0$. A deployment states a requirement $g_A$ for each condition it intends to serve and asks for $\mathrm{GAR}_A(P)\ge g_A$. Usability is measured on a single attempt because a retry is an attempt that the same budget also grants an impostor.

Algorithm~\ref{alg:select} turns a risk target and these requirements into a policy. Its one non-standard requirement is the input: every trial must carry the scores of \emph{all} subsets computed on the same samples, because the risk of a flexible policy depends on the joint behavior of its entries. Scores are computed from the fixed-point encodings, because that is the decision the servers compute (Section~\ref{ssec:design-range}), and rescaled by $2^{-2f}$, so that thresholds range over the grid $2^{-2f}\mathbb Z\cap[-1,1]$.

\begin{algorithm}[!t]
\caption{Policy selection over a finite search family}
\label{alg:select}
\small
\begin{algorithmic}[1]
\Require search and certification trials with joint encoded scores $(s_S)_{S\in\mathcal S}$ rescaled by $2^{-2f}$, labels, and identities; risk target $\varepsilon$; impostor class $\mathsf A$ with its attempt schedule; budgets $\mathcal B$; conditions $\mathcal U$ with requirements $g_A$ and weights $w_A$; presentation rule $r_{\mathcal A}$; allocation grid $\mathcal W$
\Ensure a certified policy; \textsc{Not-certified}; or \textsc{None-found} with the violated constraints
\ForAll{$\mathcal A\neq\emptyset$ such that $r_{\mathcal A}(A)$ exists wherever $g_A>0$; $\boldsymbol\omega\in\mathcal W$; $B\in\mathcal B$}
  \State for each breakpoint $t$, set $\tau_S(t)$ to the smallest threshold on the grid with $\widehat{\mathrm{FMR}}_S\le t\,\omega_S$, for $S\in\mathcal A$; if none exists, skip this $\mathcal A$
  \State $t^*\gets$ largest breakpoint with $\mathrm{UCB}\big[\hat\rho_B(\mathcal A,\boldsymbol\tau(t);\mathsf A)\big]\le\varepsilon$; if none, record \emph{risk} as violated and continue
  \State candidate if $\mathrm{LCB}\big[\widehat{\mathrm{GAR}}_A\big]\ge g_A$ for all $A\in\mathcal U$; otherwise record the violated $A$
\EndFor
\State $P^\star\gets$ candidate with the largest $\sum_A w_A\widehat{\mathrm{GAR}}_A$
\State if no candidate exists, \Return \textsc{None-found} with the recorded violations
\State re-estimate the risk and usability of $P^\star$ alone on the certification trials; \Return $P^\star$ if both bounds hold there, else \Return \textsc{Not-certified} with the violated bounds
\end{algorithmic}
\end{algorithm}

The search family consists of the allowed families that serve every required condition ($127$ at most when $|\mathcal M|=3$), a finite grid $\mathcal W$ of risk allocations across entries, the finite set of score breakpoints, and a finite set of budgets; an entry whose target cannot be met with a threshold in the domain of $\mathcal F_{\mathsf{Auth}}$ is excluded by its policy bit. Lowering thresholds can only enlarge acceptance sets, so for a fixed attempt schedule the risk is non-decreasing in $t$. Risk is checked on an upper confidence bound and usability on lower ones; because many candidates are examined on the search trials, their bounds are not certificates, and only the selected policy is re-estimated, with simultaneous bounds, on certification identities it has not seen. \textsc{None-found} means only that no searched candidate satisfied all constraints on the search trials, and \textsc{Not-certified} that the selected candidate failed on the certification trials; neither proves that no policy exists. The procedure is exhaustive search, not an optimization algorithm.
Because every account carries all $K$ slots, the servers' work for an admitted attempt is the same whichever subsets the policy allows, so the choice can rest on risk and usability alone.

%% file: sections/06-building-blocks.tex
\section{Building Blocks}
\label{sec:blocks}

After the policy has been determined by $\mathcal{SP}$, privacy-preserving authentication reduces to three core operations on secret-shared data: computing inner products to obtain the matching scores of all slots and select the slot corresponding to the presented modality subset; comparing the resulting score with zero to evaluate the threshold condition; and verifiably releasing a single decision bit to $\mathcal{SP}$. This paper extends our conference version~\cite{wang2025flame} while preserving the same underlying principles for secure protocols. 

\subsection{Authenticated Inner Product}
\label{ssec:blocks-ip}

$\Pi_{\mathsf{IP}}$ (Figure~\ref{fig:ip}) takes an authenticated vector $X\in\mathbb Z_{2^{\ell+\sigma}}^n$, held as masked shares of $X$ and of $\alpha X$, and a masked vector $Y$, and outputs the authenticated masked share $[\![\langle X,Y\rangle]\!]$. Instead of opening each of the $n$ products and summing, each server sums its local terms and opens a single element per output. The MAC result is computed from the MAC of $X$ with a second set of triples that shares its $B$ component with the first, so authentication does not double the triples for $Y$.

\begin{figure}[!t]
\small
\begin{mybox}{Protocol $\Pi_{\mathsf{IP}}\big([\![X]\!],(\Delta_Y,\langle\lambda_Y\rangle)\big)$}
\textbf{Preprocessing} (from $\mathcal F_{\mathsf{Prep}}$): shares of $A_1,A_2,B\in\mathbb Z_{2^{\ell+\sigma}}^n$, $C_1=A_1\odot B$, $C_2=A_2\odot B$; output masks $\langle\lambda_z\rangle,\langle\lambda_{\alpha z}\rangle$; public offsets $\delta_X=A_1-\lambda_X$, $\delta_{\alpha X}=A_2-\lambda_{\alpha X}$, $\delta_Y=B-\lambda_Y$.\\[2pt]
\textbf{Online}, for $j\in\{0,1\}$, with $\bar X=\Delta_X+\delta_X$, $\bar X'=\Delta_{\alpha X}+\delta_{\alpha X}$, $\bar Y=\Delta_Y+\delta_Y$:
\begin{enumerate}[leftmargin=*,topsep=0pt]
\item $\langle\Delta_z\rangle_j\gets\sum_{i}\big(j\,\bar X_i\bar Y_i-\langle A_{1,i}\rangle_j\bar Y_i-\bar X_i\langle B_i\rangle_j+\langle C_{1,i}\rangle_j\big)+\langle\lambda_z\rangle_j$.
\item $\langle\Delta_{\alpha z}\rangle_j\gets\sum_{i}\big(j\,\bar X'_i\bar Y_i-\langle A_{2,i}\rangle_j\bar Y_i-\bar X'_i\langle B_i\rangle_j+\langle C_{2,i}\rangle_j\big)+\langle\lambda_{\alpha z}\rangle_j$.
\item Exchange shares; both servers obtain $\Delta_z,\Delta_{\alpha z}$ and record them for the MAC check.
\end{enumerate}
\textbf{Output:} $[\![z]\!]=\big((\Delta_z,\langle\lambda_z\rangle),(\Delta_{\alpha z},\langle\lambda_{\alpha z}\rangle)\big)$.
\end{mybox}
\caption{Authenticated inner product with sum-then-open. All arithmetic is in $\mathbb Z_{2^{\ell+\sigma}}$; $\odot$ is the element-wise product.}
\label{fig:ip}
\end{figure}

\begin{lemma}[Correctness of $\Pi_{\mathsf{IP}}$]
\label{lem:ip}
In an honest execution with correctly authenticated $X$, a masked $Y$ held identically by both servers, and preprocessing that satisfies $C_1=A_1\odot B$, $C_2=A_2\odot B$ and the offsets are as in Figure~\ref{fig:ip}, then $\Delta_z-\lambda_z=\langle X,Y\rangle$ and $\Delta_{\alpha z}-\lambda_{\alpha z}=\alpha\langle X,Y\rangle$ in $\mathbb Z_{2^{\ell+\sigma}}$.
\end{lemma}

\begin{proof}
With $\bar X=\Delta_X+\delta_X=X+A_1$ and $\bar Y=\Delta_Y+\delta_Y=Y+B$, the two servers' shares from step~1 sum, for each coordinate $i$, to $\bar X_i\bar Y_i-A_{1,i}\bar Y_i-\bar X_iB_i+C_{1,i}=(\bar X_i-A_{1,i})(\bar Y_i-B_i)=X_iY_i$, using $C_{1,i}=A_{1,i}B_i$; hence $\Delta_z=\langle X,Y\rangle+\lambda_z$. The same computation with $\bar X'=\alpha X+A_2$ and $C_2=A_2\odot B$ gives $\Delta_{\alpha z}=\alpha\langle X,Y\rangle+\lambda_{\alpha z}$. All identities hold in $\mathbb Z_{2^{\ell+\sigma}}$.
\end{proof}

Each server sends two ring elements in one round regardless of $n$; the work and the preprocessing are linear in $n$. The integrity of the output rests on the MAC of $X$ alone, which suffices because every $Y$ that \sysname uses, is an input of the honest client application that both servers receive and compare; a server that deviates locally, including by using a different $Y$, only changes the shares it sends, which the MAC check of Section~\ref{ssec:blocks-release} detects. Because $\lambda_z$ is fixed in preprocessing, the dealer can compute the offsets of later gates, so outputs of $\Pi_{\mathsf{IP}}$ feed directly into further inner products. \sysname calls $\Pi_{\mathsf{IP}}$ $K+2$ times per attempt: $K$ times to score the probe and twice to select with the indicator $\boldsymbol\chi$.

Public-constant multiplication, which the protocol uses to place thresholds and constants on the score scale, must scale the mask together with the masked value: $c\cdot(\Delta_x,\langle\lambda_x\rangle)=(c\Delta_x,c\langle\lambda_x\rangle)$. The conference version kept $\lambda_{cx}=\lambda_x$, which reconstructs $cx+(c-1)\lambda_x$ instead of $cx$.

\subsection{Comparison on the Ring}
\label{ssec:blocks-cmp}

The decision needs $\mathbf 1[z\ge0]$ for a margin $z$ held as $(\Delta_z,\langle\lambda_z\rangle)$. The conference version computed it as $1-\mathbf 1[\Delta_z<\lambda_z]$ with a single comparison FSS key, arguing that $\mathbf 1[z+\lambda_z<\lambda_z]=\mathbf 1[z<0]$. The equality holds over the integers but not in $\mathbb Z_{2^\ell}$, where $z+\lambda_z$ wraps around. For $\ell=3$, $z=1$, and $\lambda_z=7$, $\Delta_z=0<7$, so the formula outputs $0$ although $z\ge0$. For an input $z$ in the signed range, the formula fails for exactly $|z|$ of the $2^\ell$ masks, so it returns the wrong bit with probability $|z|/2^\ell$, and on a quarter of all input--mask pairs. The error is rare for margins that are small compared to the ring, but the conference version claimed, and purported to prove, exact correctness.

The fix is to compare on the ring with a gate designed for masked inputs. A non-negative margin is one whose residue lies in the interval $[0,2^{\ell-1}-1]$, so $\mathbf 1[z\ge0]$ is an instance of the interval-containment gate of Boyle et al.~\cite{boyle2021function}. With $D=\Delta_z\bmod2^\ell$, $r=\lambda_z\bmod2^\ell$, $p'=r$, and $q'=(2^{\ell-1}-1+r)\bmod2^\ell$,
\begin{equation}
\label{eq:ic}
\mathbf 1[z\ge0]\;=\;\mathbf 1[D\le q']-\mathbf 1[D<p']+\mathbf 1[p'>q'].
\end{equation}
The two indicator terms on the right are DCF evaluations at points that the dealer knows, because it chose $r$; the last term is a constant the dealer can share. One case needs care for privacy rather than correctness: when $r=2^{\ell-1}$, $q'=2^\ell-1$ and the first term is identically one. Replacing its key by a constant would reveal $r$, and with it the compared value; we therefore evaluate both DCFs over an $(\ell+1)$-bit domain, on which $q'+1$ is an ordinary point and the keys look alike for every mask. Using the payload $(1,\alpha)\in\mathbb Z_{2^{\ell+\sigma}}^2$ for both DCF keys and sharing $\mathbf 1[p'>q']\cdot(1,\alpha)$, the servers obtain additive shares of a bit and of its MAC, add fresh output masks, and open---one round, as with the conference version's comparison. Figure~\ref{fig:cmp} gives the protocol.

\begin{figure}[!t]
\small
\begin{mybox}{Protocol $\Pi_{\mathsf{Cmp}}\big((\Delta_z,\langle\lambda_z\rangle)\big)$}
\textbf{Preprocessing:} with $r=\lambda_z\bmod2^\ell$, $p'=r$, $q'=(2^{\ell-1}-1+r)\bmod2^\ell$, the dealer generates DCF keys $(k^{\mathrm{hi}}_0,k^{\mathrm{hi}}_1)$ for $f^{<}_{q'+1,(1,\alpha)}$ and $(k^{\mathrm{lo}}_0,k^{\mathrm{lo}}_1)$ for $f^{<}_{p',(1,\alpha)}$ over the $(\ell+1)$-bit domain, so that $q'+1\le2^\ell$ is always representable and the keys have the same format for every mask; shares of $c\cdot(1,\alpha)$ with $c=\mathbf 1[p'>q']$; output masks $\langle\lambda_{b}\rangle,\langle\lambda_{\alpha b}\rangle$.\\[2pt]
\textbf{Online}, for $j\in\{0,1\}$:
\begin{enumerate}[leftmargin=*,topsep=0pt]
\item $D\gets\Delta_z\bmod2^\ell$, read as an $(\ell+1)$-bit value with leading zero.
\item $(\langle b\rangle_j,\langle\alpha b\rangle_j)\gets\mathsf{Eval}(j,k^{\mathrm{hi}}_j,D)-\mathsf{Eval}(j,k^{\mathrm{lo}}_j,D)+\langle c(1,\alpha)\rangle_j$.
\item Open $\Delta_b=b+\lambda_b$ and $\Delta_{\alpha b}=\alpha b+\lambda_{\alpha b}$; record them for the MAC check.
\end{enumerate}
\textbf{Output:} $[\![b]\!]$ with $b=\mathbf 1[z\ge0]$.
\end{mybox}
\caption{Comparison with zero on $\mathbb Z_{2^\ell}$ with authenticated payload. The input must satisfy $-2^{\ell-1}\le z<2^{\ell-1}$.}
\label{fig:cmp}
\end{figure}

\begin{lemma}[Correctness of $\Pi_{\mathsf{Cmp}}$]
\label{lem:cmp}
If $-2^{\ell-1}\le z<2^{\ell-1}$ and the preprocessing is well formed, then for every mask $\lambda_z$, $\Delta_b-\lambda_b=\mathbf 1[z\ge0]$ and $\Delta_{\alpha b}-\lambda_{\alpha b}=\alpha\,\mathbf 1[z\ge0]$ in $\mathbb Z_{2^{\ell+\sigma}}$.
\end{lemma}

\begin{proof}
Reduction modulo $2^\ell$ is a ring homomorphism, so $D=(\bar z+r)\bmod2^\ell$ with $\bar z=z\bmod2^\ell$. Since $-2^{\ell-1}\le z<2^{\ell-1}$, $z\ge0$ {if} $\bar z\in[0,2^{\ell-1}-1]$, and adding $r$ maps this interval to the cyclic interval from $p'$ to $q'$. Membership of $D$ in it is $\mathbf 1[D\le q']-\mathbf 1[D<p']$ if $p'\le q'$, and $1-\mathbf 1[D<p']+\mathbf 1[D\le q']$ if the interval wraps; both equal the right-hand side of Equation~\eqref{eq:ic}. The DCF keys output shares of $\mathbf 1[D<\gamma]\cdot(1,\alpha)$ for their points $\gamma$ and the dealer shares $\mathbf 1[p'>q']\cdot(1,\alpha)$, so by linearity the servers hold shares of $b\cdot(1,\alpha)$ with $b=\mathbf 1[z\ge0]$, and the output masks give $\Delta_b-\lambda_b=b$ and $\Delta_{\alpha b}-\lambda_{\alpha b}=\alpha b$.
\end{proof}

The input range is a real precondition, not a formality: a margin outside $[-2^{\ell-1},2^{\ell-1})$ has a residue that the gate classifies by its wrapped value. Section~\ref{ssec:design-range} chooses $\ell$ so that every margin the protocol compares lies in range. 

\subsection{Verification and Release}
\label{ssec:blocks-release}

Every value $x$ opened during an attempt is opened in masked form, as $\Delta_x$ together with $\Delta_{\alpha x}$; the relation that must hold, $\Delta_{\alpha x}-\lambda_{\alpha x}=\alpha(\Delta_x-\lambda_x)$, involves the masks and the key. The dealer therefore also provides shares of $\alpha\lambda_x$ for each value that will be opened, so that the servers can form shares of the check value
\[
c_x=\Delta_{\alpha x}-\langle\lambda_{\alpha x}\rangle-\langle\alpha\rangle\Delta_x+\langle\alpha\lambda_x\rangle,
\]
which is zero in $\mathbb Z_{2^{\ell+\sigma}}$ exactly when the opened pair is consistent. $\Pi_{\mathsf{MAC}}$ verifies all of them at once, following the batch check of SPDZ$_{2^k}$~\cite{cramer2018spd}: after every value of the attempt has been opened, the servers draw public coefficients $\gamma_x\in\mathbb Z_{2^\sigma}$ with a coin-tossing functionality, compute shares of $w=\sum_x\gamma_xc_x$, commit to their shares, open the commitments, and accept iff $w=0$. Because each $\lambda_x$ is uniform over the whole lifted ring, the opened $\Delta_x$ already hides the upper bits of $x$, which SPDZ$_{2^k}$ masks with an extra random term. An error $e$ on an opened $\Delta_x$ and $e'$ on $\Delta_{\alpha x}$ give the check value $c_x=e'-\alpha e$, as in SPDZ$_{2^k}$, whose analysis therefore applies: an error that changes some opened value modulo $2^\ell$ survives the check with probability at most $\epsilon_{\mathrm{mac}}(\sigma)=2^{-\sigma+\log(\sigma+1)}$, and one that changes only the upper $\sigma$ bits affects no value modulo $2^\ell$ and hence no decision. The outcome of a check depends on $\alpha$, so after a failed check the honest server stops and $\mathcal{SP}$, treating the failure as evidence of a deviating server, re-keys the stored state before service resumes.

One rule governs when checks run: an opened value must be verified before it is used as a public operand in a product with secret data, because an error in such an operand is multiplied by the secret and a failing check would reveal it. Only if the final check passes do the servers release the decision, to $\mathcal{SP}$ alone: each sends $(\mathsf{sid},\Delta_b,\Delta_{\alpha b})$, and $\mathcal{SP}$, which as the dealer knows $\lambda_b$, $\lambda_{\alpha b}$, and $\alpha$, checks that both servers sent the same values, computes $b=\Delta_b-\lambda_b$ and $t=\Delta_{\alpha b}-\lambda_{\alpha b}$ in $\mathbb Z_{2^{\ell+\sigma}}$, and accepts $b$ only if $b\in\{0,1\}$ and $t=\alpha b$; otherwise it treats the attempt as aborted. The servers hold only $\Delta_b$, masked by $\lambda_b$, and never see $b$.


%% file: sections/07-design.tex
\section{The Design of \sysname}
\label{sec:design}

This section assembles the building blocks into a protocol for $\mathcal F_{\mathsf{Auth}}$. Its organizing constraint is the one that Section~\ref{sec:model} placed on the functionality: the subset is an input. The servers must therefore run the same computation for every subset and every policy, so that nothing in their view depends on either.

\subsection{Enrollment}
\label{ssec:design-enroll}

For every subset $k$ that the user enrolls, the device computes a template $T_{u,k}$ from enrollment samples, normalized and encoded as the probe is (Section~\ref{ssec:model-system}); every other slot gets a random unit vector as a dummy template. It sends all $K$ masked templates $\Delta_{T_{u,k}}=T_{u,k}+\lambda_{T_{u,k}}$, with masks received from $\mathcal{SP}$, to both servers, which compare digests of what they received; the messages have the same form whichever subsets the user enrolled. The MAC component requires no interaction with the client: with shares of $\alpha\lambda_{T_{u,k}}$ from the dealer, the servers compute shares of $\alpha T_{u,k}=\langle\alpha\rangle\Delta_{T_{u,k}}-\langle\alpha\lambda_{T_{u,k}}\rangle$ locally, add a fresh mask, and open it under an enrollment run of the MAC check. $\mathcal{SP}$ sets the thresholds $\tau_{u,k}$ and policy bits $\pi_{u,k}$, with $\pi_{u,k}=0$ for dummy slots, and, knowing all masks and the key, hands the servers authenticated masked shares of $\hat\tau_{u,k}$ and $\pi_{u,k}$ directly. The servers therefore cannot tell enrolled slots from dummy ones or which subsets an account allows, and changing a policy replaces only these shares.

\subsection{Per-Attempt Preprocessing}
\label{ssec:design-prep}

For each admitted attempt, $\mathcal{SP}$ prepares under a fresh preprocessing identifier, independently of any biometric data: masks for the probe $e$ and the indicator $\boldsymbol\chi$, which it sends to the device; for the $K$ scoring inner products, one vector $B\in\mathbb Z_{2^{\ell+\sigma}}^d$ shared by all slots, since they all multiply the same probe, and fresh $A_1^{(k)},A_2^{(k)}$ with the corresponding products and offsets; for the two selections, one vector $B'\in\mathbb Z_{2^{\ell+\sigma}}^K$; the DCF keys and constants of one $\Pi_{\mathsf{Cmp}}$; and, for every value to be opened, shares of the product of the key with its mask. Sharing $B$ is the matrix form of a Beaver triple: each template is masked afresh by its own $A^{(k)}$, so templates opened in one attempt are re-randomized in the next. The material is linear in $Kd$ (Section~\ref{ssec:analysis-cost}).

\subsection{Authentication Protocol}
\label{ssec:design-auth}

{Figure~\ref{fig:auth} presents the authentication protocol
$\Pi_{\mathsf{Auth}}$.} Its correctness relies on the range
conditions required by the folding and comparison steps. Specifically, \phantomsection\label{ssec:design-range}Lemma~\ref{lem:cmp} requires $-2^{\ell-1}\le z'<2^{\ell-1}$ and folding requires $|z|<M$: the encoded unit vectors have norm at most $2^f+\sqrt d/2\le2^{f+1}$ whenever $\sqrt d\le2^{f+1}$, so by the Cauchy--Schwarz inequality $|s_k|\le2^{2f+2}$, and thresholds in $[-1,1]$ give $|z_k|<2^{2f+3}=M$ and $-2^{2f+4}<z'<2^{2f+3}$; choosing $\ell\ge2f+5$ therefore places every compared value in range for every input that satisfies the preconditions of $\mathcal F_{\mathsf{Auth}}$.
With these ranges every value of an honest execution is an exact integer: by Lemma~\ref{lem:ip}, $s_k=\langle e,T_{u,k}\rangle$; the selections give $z=s_{k^\star}-\hat\tau_{u,k^\star}$ and $a=\pi_{u,k^\star}$ for the presented slot $k^\star$; folding leaves $z'=z$ when $a=1$ and gives $z'=z-M<0$ when $a=0$; and Lemma~\ref{lem:cmp} returns $b=\mathbf 1[z'\ge0]=\pi_{u,k^\star}\land\mathbf 1[s_{k^\star}\ge\hat\tau_{u,k^\star}]$, the decision of $\mathcal F_{\mathsf{Auth}}$ on the encoded inputs.

\begin{figure}[!t]
\small
\begin{mybox}{Protocol $\Pi_{\mathsf{Auth}}$ for account $u$}
\begin{enumerate}[leftmargin=*,topsep=0pt]
\item \textbf{Admission} ($\mathcal{SP}$). If $u$ has used $B$ attempts in the current window, answer \textsf{locked}. Otherwise record the attempt, then send an authenticated admission record---the admission metadata of $\mathcal F_{\mathsf{Auth}}$ and a preprocessing identifier--- with the attempt's preprocessing to $\mathcal P_0,\mathcal P_1$, and the input masks $\lambda_e,\lambda_{\boldsymbol\chi}$ to $\mathcal C$.
\item \textbf{Input} ($\mathcal C$). Send $(\mathsf{sid},\Delta_e=e+\lambda_e,\ \Delta_{\boldsymbol\chi}=\boldsymbol\chi+\lambda_{\boldsymbol\chi})$ to both servers. The servers exchange a digest of what they received and abort on mismatch.
\item \textbf{Score.} For $k\in[K]$: $[\![s_k]\!]\gets\Pi_{\mathsf{IP}}([\![T_{u,k}]\!],(\Delta_e,\langle\lambda_e\rangle))$; then $[\![z_k]\!]\gets[\![s_k]\!]-[\![\hat\tau_{u,k}]\!]$.
\item \textbf{Verify scores.} Run $\Pi_{\mathsf{MAC}}$ on the $K$ score openings of step~3; on failure, abort and stop using the key.
\item \textbf{Select.} $[\![z]\!]\gets\Pi_{\mathsf{IP}}([\![(z_k)_{k}]\!],(\Delta_{\boldsymbol\chi},\langle\lambda_{\boldsymbol\chi}\rangle)) $ and $[\![a]\!]\gets\Pi_{\mathsf{IP}}([\![(\pi_{u,k})_{k}]\!],(\Delta_{\boldsymbol\chi},\langle\lambda_{\boldsymbol\chi}\rangle))$.
\item \textbf{Fold.} $[\![z']\!]\gets[\![z]\!]+M\cdot([\![a]\!]-1)$ with the public constant $M=2^{2f+3}$.
\item \textbf{Compare.} $[\![b]\!]\gets\Pi_{\mathsf{Cmp}}([\![z']\!])$.
\item \textbf{Verify.} Run $\Pi_{\mathsf{MAC}}$ on every value opened in steps~5--7; on failure, abort and stop using the key.
\item \textbf{Release.} Each server sends $(\mathsf{sid},\Delta_b,\Delta_{\alpha b})$ to $\mathcal{SP}$, which accepts $b$ only if both messages agree, $b\in\{0,1\}$, and the MAC verifies.
\end{enumerate}
\end{mybox}
\caption{{Authentication in \sysname.} }
\label{fig:auth}
\end{figure}

\subsection{Design Rationale {and Online Optimization}}
\label{ssec:design-rationale}


Four choices in $\Pi_{\mathsf{Auth}}$ are deliberate. \emph{Score first, select second:} selecting the presented template coordinate by coordinate would open about $2d+6$ elements, whereas scoring every slot first opens $2K+4$; we score first because $K\ll d$ ($K=7$ against $d=512$). \emph{Fold the policy bit:} since $a\in\{0,1\}$ and $|z|<M$, adding $M(a-1)$ before the comparison yields the same decision as multiplying $a$ with its result, without one more authenticated multiplication and round. \emph{Verify scores before selection:} an opened value becomes a public operand of the next inner product, where the honest server multiplies it with secret data; if a corrupted server shifted the opened MAC component of one score by one, the selection product would multiply that error by the slot's indicator bit, and a later batch check would reveal the indicator even though the attempt aborts. Verifying the scores first means every check covers only errors that the corrupted server introduced directly, whose check value $e'-\alpha e$ does not depend on any secret input, at the price of one extra check (two rounds). \emph{Verify before release and bind attempts to records:} a valid MAC on the released bit does not certify the values it was derived from; running $\Pi_{\mathsf{MAC}}$ over every opened value before release, binding each attempt to an admission record and a preprocessing identifier used once, and having $\mathcal{SP}$ accept one result per session ensure that a corrupted server can cause an abort but cannot have a biased decision accepted, replay an old one, or rerun an aborted attempt with the same masks. 

{Beyond these protocol-level choices, we reduce the online cost of $\Pi_{\mathsf{Auth}}$ by moving selector validation before the online computation using {policy-certified selector authorization (PCSA)}. During {preprocessing}, {$\mathcal{SP}$} signs authorizations for all $K$ valid one-hot choices, binding each to the {policy input} and a {one-time preprocessing identifier}. During authentication preparation, the client obtains exactly one authorization through {malicious-secure $1$-out-of-$K$ oblivious transfer}, binding its selected one-hot vector, {$\Delta_{\boldsymbol\chi}$}, and the preprocessing identifier. Before declaring the inputs ready, each server verifies the PCSA therefore reduces online cost while preserving the authenticated decision semantics.}

\subsection{Extension to \texorpdfstring{$1{:}N$}{1:N} Identification}
\label{ssec:design-1n}

{We next extend the optimized $1{:}1$ authentication design to identification without a claimed identity.} $\mathcal F_{\mathsf{Ident}}$ outputs to $\mathcal{SP}$ the account $u^\star=\arg\max_u z'_u$, with ties broken by index, if $z'_{u^\star}\ge0$, and $\bot$ otherwise, where $z'_u$ is the folded margin of account $u$ for the presented subset. Computing all $z'_u$ costs $N$ times steps~3--5; a tournament of $\Pi_{\mathsf{Cmp}}$ calls on differences of margins, each followed by an authenticated selection of the larger margin and its index, finds the maximum in $\lceil\log_2N\rceil$ levels, a last comparison decides acceptance, and $\ell\ge2f+6$ is needed because a difference of two margins can be twice a margin. Work and preprocessing are linear, shown in Table.~\ref{tab:cost}. The extension serves to measure how the cost scales (Section~\ref{ssec:eval-rq4}); its security is not covered by Theorem~\ref{thm:main}.

%% file: sections/08-analysis.tex
\section{Theoretical Analysis}
\label{sec:analysis}

\subsection{Security}
\label{ssec:analysis-security}

Security is defined by the standard simulation paradigm with abort for an adversary~\cite{canetti2000security,lindell2017simulate}; $\epsilon_{\mathrm{mac}}(\sigma)=2^{-\sigma+\log(\sigma+1)}$ is the error of the batch MAC check (Section~\ref{ssec:blocks-release}).

\begin{theorem}
\label{thm:main}
Assume the $\mathcal F_{\mathsf{Prep}}$-hybrid model with a coin-tossing functionality, 
an honest client application, a collision-resistant digest, and that the key is replaced after any failed MAC check. Then $\Pi_{\mathsf{Auth}}$ securely computes $\mathcal F_{\mathsf{Auth}}$ with abort against a probabilistic polynomial-time adversary that statically corrupts at most one of $\mathcal P_0,\mathcal P_1$. For each attempt, the distinguishing advantage is at most $\epsilon_{\mathrm{mac}}(\sigma)+\mathrm{negl}(\kappa)$, where $\epsilon_{\mathrm{mac}}(\sigma)$ is the statistical failure probability of the MAC verification and the negligible term arises from the computational security of the DCFs, the commitment scheme, and the digest.
\end{theorem}

\begin{proof}[Proof idea]
We outline the argument for a corrupted $\mathcal P_1$. The simulator generates all preprocessing itself, including $\alpha$, and runs the honest parties' code on dummy inputs. Every public value that depends on an input is masked by a fresh, uniform mask of which the corrupted server holds only a share, so the view is distributed identically for any inputs once the corrupted server's DCF keys are replaced by keys for an independent mask, which costs $\mathrm{negl}(\kappa)$. Knowing every share, the simulator sees each deviation and runs the MAC and release checks exactly as the honest parties would, so both worlds abort in the same executions; because every opened value is verified before it multiplies a secret, the check values depend only on the corrupted server's own errors and on $\alpha$, never on the inputs. They differ only in one event: a deviation that changes an opened value modulo $2^\ell$ and still passes the check. Since each opened pair contributes the check value $e'-\alpha e$, as in the SPDZ$_{2^k}$ batch check, that event has probability at most $\epsilon_{\mathrm{mac}}(\sigma)$. Deviations confined to the upper $\sigma$ bits change no value modulo $2^\ell$; without deviations, $\mathcal{SP}$ obtains the functionality's decision (Section~\ref{ssec:design-range}).
\end{proof}

The theorem is a standalone statement and composes sequentially~\cite{canetti2000security}
. The key-replacement assumption is essential and is a deployment obligation: whether a check passes depends on $\alpha$, so a server allowed to trigger checks repeatedly under one key could learn bits of it. 

\begin{corollary}[Subset privacy]
\label{cor:subset}
Under the conditions of Theorem~\ref{thm:main}, for fixed public metadata and the leakage allowed by $\mathcal F_{\mathsf{Auth}}$, the view of a corrupted server can be simulated without the client's embedding, the presented subset, the policy, or the decision. The view of $\mathcal{SP}$ can be simulated from its own inputs and randomness and the outputs $\mathcal F_{\mathsf{Auth}}$ gives it.
\end{corollary}

\begin{proof} 
The first claim follows directly from Theorem~\ref{thm:main}. In the simulator constructed there, every opened value observed by a corrupted server is masked by fresh randomness, so the simulator never requires the client's embedding, the presented subset, the policy, or the authentication outcome. For the second claim, $\mathcal{SP}$ receives only $\Delta_b$ and $\Delta_{\alpha b}$, both of which are determined by its own randomness and by the output prescribed by $\mathcal F_{\mathsf{Auth}}$. Hence its view can be simulated from this information alone. 
\end{proof}


\begin{remark}[Subset privacy is relative to the decision]
\label{rem:tv}
Consider two authentication entries $S_0$ and $S_1$. Let $b \in \{0,1\} $ denote the authentication decision, where $b=1$ means acceptance and $b=0$ means rejection. Let $V_0$ and $V_1$ denote the views of a party that learns the decision $b$ when the same user authenticates through $S_0$ and $S_1$, respectively. Suppose that the two entries accept the user with different probabilities: \[ \Pr[b=1 \mid S_0] \neq \Pr[b=1 \mid S_1]. \] Then observing the authentication decision provides statistical evidence about which entry was used. Formally, let $\mathrm{TV}(V_0,V_1) $ be the total variation distance between the two view distributions, i.e., the maximum distinguishing advantage achievable by any observer. Since the decision $b$ is part of the view, we have \[ \mathrm{TV}(V_0,V_1) \ge \Big| \Pr[b=1\mid S_0] - \Pr[b=1\mid S_1] \Big|. \] Therefore, whenever different authentication entries have different acceptance probabilities, a party that learns the authentication outcome inevitably learns some information about the presented subset. This leakage originates from the functionality output itself rather than from the protocol messages, and therefore cannot be eliminated by any protocol.

\end{remark}

\subsection{Cost}
\label{ssec:analysis-cost}

Table~\ref{tab:cost} itemizes the cost of one attempt. Counts are elements of $\mathbb Z_{2^{\ell+\sigma}}$ sent per server unless stated; ``local'' operations involve no communication.
Two readings of the table connect it to the design goals. The server-to-server online communication of a verification does not depend on $d$, as G4 requires; what grows with $d$ is the client's input, the local work, the preprocessing, and the storage
. Subset privacy costs a factor $K$ in work and preprocessing relative to scoring a single known slot, and nothing that depends on the policy.

\begin{table}[!t]
\centering
\caption{Itemized cost per admitted attempt for per server. $K$: slots; $d$: dimension; $N$: accounts; $c_{\mathrm{dcf}}$: bits of one DCF key over $(\ell+1)$-bit inputs with a two-element payload
.}
\label{tab:cost}
\small
\resizebox{\columnwidth}{!}{%
\begin{tabular}{@{}lll@{}}
\toprule
Item & $1{:}1$ (\sysname) & $1{:}N$ extension \\
\midrule
Input from client & $d+K$ & $d+K$ \\
$\Pi_{\mathsf{IP}}$ calls & $K+2$ & $N(K+2)$ \\
Openings  & $2K+5$ & $N(2K+4)+1$ \\
Arithmetic rounds & $3$ & $2+2\lceil\log_2N\rceil+1$ \\
$\Pi_{\mathsf{MAC}}$ runs & $2$  & $2$ \\
Local multiplications & $O(Kd)$ & $O(NKd)$ \\
DCF evaluations & $2$ & $2N$ \\
Preprocessing & $O(Kd)$ $+\,2c_{\mathrm{dcf}}$ & $O(NKd)$  $+\,2Nc_{\mathrm{dcf}}$ \\
\bottomrule
\end{tabular}}
\end{table}

%% file: sections/09-evaluation.tex
\section{Experimental Evaluation}
\label{sec:eval}

We investigate four research questions: (RQ1)~how multimodal authentication compares with single-modality authentication in recognition performance; (RQ2)~to what extent modality flexibility increases impostor risk, and how the risk evolves under different attacker capabilities and attempt budgets; (RQ3)~whether the secure-computation implementation faithfully reproduces the authentication decisions of the plaintext system; and (RQ4)~what computational and communication costs privacy-preserving authentication incurs.


\subsection{Setup}
\label{ssec:eval-setup}


\paragraph*{Data} We use two datasets with different provenance and report them separately. \emph{VidTIMIT}~\cite{vidtimit2009dataset,sanderson2009vidtimit} provides video and speech of 43 people recorded in three sessions, ten sentences per person (six in Session~1 and two in each of Sessions~2--3). Session~1 is used for enrollment and Sessions~2--3 for probes, yielding four probe bundles per identity and therefore $43\times(2+2)=172$ genuine trials. \emph{MULB+} combines the hand and iris images of the MULB dataset~\cite{mulb2024dataset} with face images from LFW~\cite{huang2007lfw}. We selected the identities with at least 13 distinct hand and 13 distinct iris images, obtaining 136 identities, of which a seeded 20\% (27) are reserved for model selection and the remaining 109 are evaluated. As MULB+ has no session structure, enrollment and probe sets are formed from disjoint random image subsets of each identity. Three probe bundles are constructed per evaluated identity, yielding $109\times3=327$ genuine trials. Each MULB identity is assigned the face images of a distinct LFW identity with at least six images. MULB+ identities are therefore \emph{chimeric}: the hand and iris belong to one person whereas the face belongs to another. A \emph{bundle} consists of one sample from each available modality: a video--audio pair in VidTIMIT or the corresponding probe images in MULB+. 
A \emph{trial} pairs a bundle with a claimed identity and is scored by all modality subsets. Experimental evaluations generate 172 genuine and 1,312 impostor trials for VidTIMIT, and 327 genuine and 6,804 impostor trials for MULB+.

\paragraph*{Models}
No model is trained or fine-tuned: every modality uses a frozen, publicly released extractor, and every vector is normalized to unit length. Faces (a VidTIMIT recording contributes the renormalized mean of eight evenly spaced frames) are embedded with AuraFace-v1~\cite{fal2024auraface}, a ResNet-100 trained with the ArcFace loss~\cite{deng2019arcface}; speech with ECAPA-TDNN~\cite{desplanques2020ecapa}, one vector per sentence; visible-light irises with the class token of DINOv2-L~\cite{oquab2024dinov2}; 
and hands with the palmprint network CCNet~\cite{yang2023ccnet} on a $128\times128$ gray-scale palm square~\cite{zhang2020mediapipe}. 
The other pre-declared candidates were an ArcFace ResNet-50 trained on WebFace600K and the Open-IRIS normalized iris band. 
Fusion has three candidates that all score with one inner product: equal-weight concatenation (Fu1), which equals the sum rule~\cite{kittler1998combining}; z-score weights (Fu2, $\alpha_m=1/\sigma_m$ with $\sigma_m$ the standard deviation of impostor scores of modality $m$ on selection data); and Fu1 followed by a fixed Johnson--Lindenstrauss projection to $d=512$ (Fu3)~\cite{dasgupta2003elementary}. 
A selection rule registered before any score was computed takes per modality the extractor with the lowest false non-match rate at a false match rate of 1\% and then the fusion with the lowest such rate averaged over the multimodal subsets, breaking ties by smaller dimension. It selected Fu1 ($d=D=3584$) on the 27 MULB+ development identities, and AuraFace, ECAPA-TDNN, and Fu3 ($d=512$, by the dimension tie-break) inside the search group of every VidTIMIT rotation. 


\paragraph*{Performance Metrics} For an entry $S$ with threshold $\tau$, the false match rate is $\mathrm{FMR}_S(\tau)=\Pr[s_S\ge\tau\mid\text{impostor trial}]$ and the false non-match rate is $\mathrm{FNMR}_S(\tau)=\Pr[s_S<\tau\mid\text{genuine trial}]$; the genuine acceptance rate is $\mathrm{GAR}=1-\mathrm{FNMR}$. The equal error rate (EER) is the rate at which the two errors are equal. The EER summarizes a curve over all thresholds and is not an operating point: a deployment fixes a threshold at a target FMR and reports the FNMR at that frozen threshold (FNMR@FMR). For secure computation protocols, we additionally report communication volume, communication rounds, and end-to-end latency. Communication volume denotes the total data exchanged between the two parties, communication rounds the number of sequential interactions, and latency the time from query submission to authentication decision. 

\paragraph*{Implementation and environment} We implement a prototype of \sysname{} in C++. All experiments are conducted on two servers, each equipped with
{{an Intel Xeon E5-2682 v4 processor, 32 vCPUs,
256\,GB of RAM, and a 400\,GB cloud disk}}.
{Both servers run Ubuntu 20.04.6 LTS.} Unless otherwise stated, we use $\ell=\sigma=64$, $f=16$, and $\kappa=128$. Network conditions are emulated using the Linux \texttt{tc} traffic-control utility. We evaluate three representative settings: {{LAN (RTT $=0.1$\,ms, bandwidth $=10$\,Gbps)
and WAN (RTT $=40$\,ms, bandwidth $=600$\,Mbps)}.}. Every timing result is reported as the median and the 95th percentile over at least 30 runs after warm-up. Plain models are conducted with Intel Core i7-12700F, NVIDIA RTX~3080, PyTorch 2.14 with CUDA 13.0.

\begin{table}[!t]
\centering
\caption{Per-subset behavior on the test group (RQ1).
$\tau$ is the threshold for a 1\% search FMR (cosine scale);
EER, FNMR, and realized FMR in percent.
}
\label{tab:rq1}
\small
\resizebox{\columnwidth}{!}{%
\begin{tabular}{@{}llcccccc@{}}
\toprule
\multirow{2}{*}{Data} & \multirow{2}{*}{Subset} & \multirow{2}{*}{$\tau$}
& \multirow{2}{*}{EER}
& \multicolumn{2}{c}{FNMR @ target FMR}
& realized FMR \\
\cmidrule(lr){5-6}
& & & & 1\% & 5\% & at 1\% \\
\midrule
\multirow{7}{*}{MULB+}
& F & 0.234 & 0.92 & 0.92 & 0.92 & 1.03 \\
& I & 0.979 & 8.64 & 21.10 & 10.40 & 1.09 \\
& H & 0.836 & 1.53 & 1.53 & 0.31 & 1.21 \\
& F+I & 0.591 & 0.92 & 0.92 & 0.61 & 1.00 \\
& F+H & 0.500 & 0.61 & 0.31 & 0.31 & 0.98 \\
& I+H & 0.895 & 0.31 & 0.00 & 0.00 & 1.13 \\
& F+I+H & 0.650 & 0.51 & 0.31 & 0.31 & 1.01 \\
\midrule
\multirow{3}{*}{VidTIMIT}
& F & 0.291 & 0.00 & 0.00 & 0.00 & 1.68 \\
& A & 0.382 & 0.91 & 0.58 & 0.00 & 0.84 \\
& F+A & 0.290 & 0.00 & 0.00 & 0.00 & 1.37 \\
\bottomrule
\end{tabular}}
\end{table}


\subsection{RQ1: Single- versus Multi-Modality Accuracy}
\label{ssec:eval-rq1}

\begin{table}[t]
\centering
\caption{Sensitivity of the MULB+ EER (\%) to various choices. Each row changes one choice of the primary pipeline (AuraFace, DINOv2-L, CCNet, Fu1). }
\label{tab:rq1-sens}
\small
\resizebox{\columnwidth}{!}{%
\begin{tabular}{@{}lccccccc@{}}
\toprule
Variant & F & I & H & F+I & F+H & I+H & F+I+H \\
\midrule
Primary & 0.92 & 8.64 & 1.53 & 0.92 & 0.61 & 0.31 & 0.51 \\
Three samples  & 0.92 & 9.48 & 3.67 & 0.92 & 0.50 & 2.45 & 0.31 \\
Fusion Fu2   & 0.92 & 8.64 & 1.53 & 0.92 & 0.61 & 0.12 & 0.31 \\
Fusion Fu3  & 0.92 & 8.56 & 1.59 & 1.22 & 0.92 & 0.60 & 0.61 \\
Face: ResNet-50 & 0.61 & 8.64 & 1.53 & 0.61 & 0.31 & 0.31 & 0.31 \\
Hand: DINOv2-L & 0.92 & 8.64 & 10.85 & 0.92 & 0.92 & 5.35 & 0.92 \\
Iris: Open-IRIS & 0.92 & 18.71 & 1.53 & 10.70 & 0.61 & 11.62 & 10.40 \\
\bottomrule
\end{tabular}}
\end{table}

This experiment measures the accuracy of every subset and asks whether fusing modalities beats the best single modality. Table~\ref{tab:rq1} also reports the realized FMR, since a target fixed on one group of people is not guaranteed on another; with fewer than two hundred identities, these operating points are diagnostic rather than deployment targets.
%

On MULB+, the face is the strongest single modality with an EER of 0.92\%, followed by the hand (1.53\%) and the visible-light iris (8.64\%). Adding the hand to another modality consistently reduces the EER relative to the face: F+H achieves 0.61\%, I+H 0.31\%, and F+I+H 0.51\%. Notably, the advantage of I+H is sensitive to the number of enrollment images, with performance degrading substantially when fewer enrollment samples are available.
In contrast, combining the face with the iris alone does not improve performance, as F+I has the same EER as the face (0.92\%). The FNMR values show the same pattern. At the operating point defined by the search-group threshold for a target 1\% FMR, I+H achieves 0\% FNMR, while F+H and F+I+H reduce the FNMR to 0.31\% compared with 0.92\% for the face alone. The realized FMR remains close to the target value of 1\% for all subsets. 
On VidTIMIT, the face modality completely separates genuine and impostor trials, yielding an EER and FNMR of 0\%. Speech alone achieves an EER of 0.91\% and an FNMR of 0.58\% at the target 1\% FMR. Fusion (F+A) matches the performance of the face. However, the realized FMR differs from the nominal operating point, reaching 1.68\% for the face, 0.84\% for speech, and 1.37\% for the fused system, indicating that thresholds calibrated on one group do not necessarily transfer exactly to another.


Table~\ref{tab:rq1-sens} shows that the ranking of modalities depends more on the extractors than on the fusion rule. Reducing enrollment from ten images per modality to three weakens the advantage of I+H (0.31\%\,$\rightarrow$\,2.45\%), whereas changing from Fu1 to Fu2 or Fu3 leaves the overall conclusions unchanged. In contrast, replacing CCNet with DINOv2-L degrades the hand modality substantially (1.53\%\,$\rightarrow$\,10.85\%), and replacing the whole iris image with the Open-IRIS normalized band degrades all iris-containing combinations (F+I+H: 0.51\%\,$\rightarrow$\,10.40\%). 


\begin{table}[!t]
\centering
\caption{Risk quantities characterizing modality flexibility (RQ2). Each entry at its own threshold $\tau_{1\%}$, and brackets indicate 95\% identity-bootstrap intervals. 
}
\label{tab:rq2_risk}
\small
\resizebox{\columnwidth}{!}{%
\begin{tabular}{llcccc}
\toprule
Dataset & Subset &
$R_{\mathrm{weak}}$ &
$U$ &
$\Sigma$ &
$\Delta_{\mathrm{sel}}$ \\
\midrule
\multirow{5}{*}{MULB+}
 & $\{F,I,H\}$ & 1.21 {\scriptsize[0.90, 2.44]} & 3.31 {\scriptsize[2.16, 4.65]} & 0.01 {\scriptsize[0.00, 0.06]} & 2.10 {\scriptsize[1.13, 2.62]} \\
 & \begin{tabular}[c]{@{}l@{}}$\{F,I,H,F{+}I,$\\ $F{+}H,I{+}H\}$\end{tabular} & 1.21 {\scriptsize[0.92, 2.57]} & 4.26 {\scriptsize[2.83, 5.88]} & 2.18 {\scriptsize[1.05, 3.48]} & 3.06 {\scriptsize[1.70, 3.74]} \\
 & \begin{tabular}[c]{@{}l@{}}$\{F,I,H,F{+}I,$\\ $F{+}H,I{+}H,F{+}I{+}H\}$\end{tabular}
   & 1.21 {\scriptsize[0.92, 2.57]} & 4.42 {\scriptsize[2.95, 6.09]} & 3.03 {\scriptsize[1.49, 4.67]} & 3.22 {\scriptsize[1.85, 3.92]} \\
\midrule
\multirow{2}{*}{VidTIMIT}
 & $\{F,A\}$ & 1.68 {\scriptsize[0.17, 4.93]} & 2.36 {\scriptsize[0.17, 5.95]} & 0.15 {\scriptsize[0.00, 0.72]} & 0.69 {\scriptsize[0.00, 1.74]} \\
 & $\{F,A,F{+}A\}$
   & 1.68 {\scriptsize[0.18, 4.96]} & 2.52 {\scriptsize[0.26, 6.11]} & 1.37 {\scriptsize[0.00, 4.69]} & 0.84 {\scriptsize[0.00, 2.07]} \\
\bottomrule
\end{tabular}}
\end{table}

\begin{table}[!t] \centering \caption{Selection gain $D_Q$ under different attempt budgets for the Enumerator attack (RQ2), with every subset allowed, against the fixed entry $\hat S $. $|\mathcal A|=7$ for MULB+ and $3$ for VidTIMIT. } \label{tab:rq2_budget} \small \resizebox{0.9\columnwidth}{!}{%
\begin{tabular}{lcccc} \toprule Dataset & $Q=1$ & $Q=2$ & $Q=3$ & $Q=|\mathcal A|$ \\ \midrule MULB+ & 0.00 & 1.09 {\scriptsize[0.51, 1.87]} & 2.28 {\scriptsize[1.24, 3.60]} & 3.40 {\scriptsize[2.02, 5.06]} \\
VidTIMIT & 0.00 & 0.69 {\scriptsize[0.00, 2.36]} & 0.84 {\scriptsize[0.00, 2.71]} & 0.84 {\scriptsize[0.00, 2.71]} \\ \bottomrule \end{tabular}} \end{table}

\begin{figure}[t]
\centering
\includegraphics[width=0.9\columnwidth]{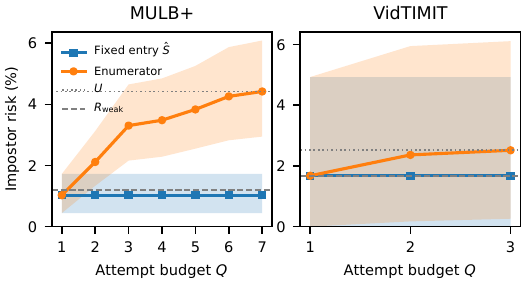}
\caption{Impostor risk against the attempt budget $Q$ (RQ2), with every subset allowed: the fixed entry $\hat S$ with bundle reuse and the enumerator, with 95\% identity-bootstrap bands; dashed and dotted lines mark $R_{\mathrm{weak}}$ and $U$.}
\label{fig:rq2-budget}
\end{figure}

\subsection{RQ2: The Risk Added by Flexibility}
\label{ssec:eval-rq2}

This experiment tests where the risk of a flexible policy falls between the weakest entry and the union of entries (Proposition~\ref{prop:ordering}), and whether an impostor who chooses entries gains anything beyond using a fixed one. We fix each subset at its own threshold for a 1\% FMR on the search group. Table~\ref{tab:rq2_risk} reports $R_{\mathrm{weak}}$, $U$, $\Sigma$, and $\Delta_{\mathrm{sel}}$, and Table~\ref{tab:rq2_budget} and Fig.~\ref{fig:rq2-budget} the gain $D_Q$ of the enumerator over the fixed entry $\hat S$, both reusing one bundle, so that the risk of the fixed entry is $p_{\hat S}$ for every $Q$. $\hat S$ is the face on both datasets.

\emph{Does flexibility raise the risk above the weakest entry?} On MULB+, yes: 301 of the 6{,}804 test impostor trials are accepted by at least one of the seven entries, so $U=4.42\%$ against $R_{\mathrm{weak}}=1.21\%$ (i.e., $3.7\times$), and the interval of $\Delta_{\mathrm{sel}}=[1.85,3.92]$ excludes zero. On VidTIMIT, $U=2.52\%$ against $R_{\mathrm{weak}}=1.68\%$ (i.e., $1.5\times$); the difference of 0.84 points comes from eleven trials and its interval reaches zero.

\emph{Does an impostor gain by switching entries, and does the gain grow with the budget?} On MULB+ the enumerator gains 1.09 points at $Q=2$, 2.28 at $Q=3$, and 3.40 at $Q=7$, each interval excluding zero, and its risk grows with every additional attempt, to $\rho_7=U$; on VidTIMIT it gains 0.69 points at $Q=2$ and 0.84 at $Q=3$, with intervals that include zero. Because $\hat S$ is not the weakest test entry on MULB+, $\rho_1=p_{\hat S}=1.03\%$ lies below $R_{\mathrm{weak}}$ and $D_7=U-p_{\hat S}=3.40$ points exceeds $\Delta_{\mathrm{sel}}=3.22$; thus $R_{\mathrm{weak}}<\rho_Q\le U$ for $Q\ge2$ (i.e., $\rho_Q= \rho_1+D_Q$), and under these thresholds opening more entries enlarges the attack surface.


\emph{Does overlap explain the growth?} Yes. The increase from $R_{\mathrm{weak}}$ to $U$ depends on how much the entries accept the same impostor trials. On MULB+, the seven entries have realized FMRs of 1.03\%, 1.09\%, 1.21\%, 1.00\%, 0.98\%, 1.13\%, and 1.01\%, which sum to $\sum_S p_S = 7.45\%$. Across the 6{,}804 impostor trials, this corresponds to $6804 * \sum_S p_S=507$ acceptance events in total. However, many of these are repeated acceptances of the same trial by multiple entries: after deduplication, only 301 distinct impostor trials are accepted by at least one entry, giving the union risk $U=301/6804=4.42\%$. The difference, $\Sigma=\sum_S p_S-U=3.03\%$, therefore measures the overlap among entry acceptance sets. Of the 301 accepted trials, 171 are accepted by exactly one entry, indicating substantial diversity across entries. On VidTIMIT, 17 of the $1312*2.52\%=33$ accepted trials are accepted by only one entry. Since all $p_S$ are close to the common target value, variation in $\Delta_{\mathrm{sel}}=\sum_S p_S-\Sigma-R_{\mathrm{weak}}$ is driven mainly by $\Sigma$: the less the acceptance sets overlap, the larger the risk added by flexibility. 

\emph{How does the choice of policy change the risk?} Restricting the allowed family keeps every entry's threshold, so $R_{\mathrm{weak}}$ is the same for every family of a dataset, whereas $U$ grows with the allowed subsets: on MULB+ from 3.31\% with the single modalities to 4.26\% with the pairs added and 4.42\% with all seven entries, and on VidTIMIT from 2.36\% to 2.52\%, with overlapping intervals between consecutive families. $\Sigma$ shows how much of what an added entry accepts is already accepted: the single modalities of MULB+ hardly overlap ($\Sigma=0.01\%$), whereas the four multimodal entries account for $6804*(1.00\%+0.98\%+1.13\%+1.01\%)=281$ acceptances but add only $301-225=76$ trials to the $6804*3.31\%=225$ that the single modalities accept, which raises $\Sigma$ to 3.03\%; on VidTIMIT, F+A accepts $1312*1.37\%=18$ trials and adds two (i.e., $1312* (2.52\%-2.36\%)$). 


\subsection{RQ3: Fidelity of Secure Execution}
\label{ssec:eval-rq3}

\begin{table*}[t]
\centering
\caption{Fidelity of execution (RQ3). Agreeing and Disagree count decisions over every test trial. 
$\max|\Delta s|$: largest difference between the floating-point and the fixed-point score; $\min|s-\tau|$: smallest distance of a score to its threshold on the cosine scale. 
}
\label{tab:rq3}
\small
\begin{tabular}{@{}llccccccc@{}}
\toprule
Data & Execution & Agree & Disagree
& $\max|\Delta s|$ & $\min|s-\tau|$
& EER & FNMR & FMR \\
\midrule
\multirow{3}{*}{MULB+}
& Floating point
& 99{,}833/99{,}834 & 1
& $2.8\times10^{-5}$
& $5.8\times10^{-6}$
& 0.47 & 0.31 & 1.01 \\

& Fixed point
& 99{,}834/99{,}834 & 0
& 0
& $1.9\times10^{-6}$
& 0.47 & 0.31 & 1.01 \\

& \sysname
& 99{,}834/99{,}834 & 0
& 0
& $1.9\times10^{-6}$
& 0.47 & 0.31 & 1.01 \\

\midrule
\multirow{3}{*}{VidTIMIT}
& Floating point
& 8{,}904/8{,}904 & 0
& $2.1\times10^{-5}$
& $4.4\times10^{-6}$
& 0.00 & 0.00 & 1.37 \\

& Fixed point
& 8{,}904/8{,}904 & 0
& 0
& $1.2\times10^{-5}$
& 0.00 & 0.00 & 1.37 \\

& \sysname
& 8{,}904/8{,}904 & 0
& 0
& $1.2\times10^{-5}$
& 0.00 & 0.00 & 1.37 \\

\bottomrule
\end{tabular}
\end{table*}
\begin{table}[!t]
\centering
\caption{Plaintext cost per operation (RQ4), batch size one, median and 95th percentile of 50 runs.
}
\small
\resizebox{\columnwidth}{!}{%
\begin{tabular}{@{}llrr@{}}
\toprule
Data & Operation & Median (ms) & P95 (ms) \\
\midrule
\multirow{6}{*}{MULB+}
& Face extraction (AuraFace, 1 image) & 10.0 & 13.9 \\
& Iris extraction (DINOv2-L, 1 image) & 29.1 & 79.2 \\
& Hand extraction (CCNet, 1 image) & 68.0 & 128.8 \\
& F+I+H extraction + fusion + encoding & 117.3 & 192.2 \\
& Fusion + encoding alone ($d=3584$) & 0.03 & 0.03 \\
& Matching, one attempt ($K=7$, $d=3584$) & 0.03 & 0.03 \\
\midrule
\multirow{5}{*}{VidTIMIT}
& Face extraction (AuraFace, 8 frames) & 86.7 & 89.1 \\
& Speech extraction (ECAPA-TDNN, 1 sentence) & 6.8 & 7.6 \\
& F+A extraction + fusion + encoding & 95.5 & 97.1 \\
& Fusion + encoding alone ($d=512$) & 0.07 & 0.07 \\
& Matching, one attempt ($K=3$, $d=512$) & 0.01 & 0.01 \\
\bottomrule
\end{tabular}}
\label{tab:rq4-plain}
\end{table}

\begin{table}[!t]
\centering
\caption{Breakdown of secure computation costs per authentication attempt (in ms).}
\label{tab:rq3_part}
\small
\resizebox{\columnwidth}{!}{%
\begin{tabular}{@{}lccccccc@{}}
\toprule
\multirow{2}{*}{Dataset}
& \multicolumn{5}{c}{Server Online}
& \multicolumn{2}{c}{$\mathcal{SP}$} \\
\cmidrule(lr){2-6}
\cmidrule(l){7-8}
& Scoring & Selection & Comparison & Check
& Total & Release & Verification \\
\midrule
MULB+
& 2.65 & 0.35 & 0.99 & 2.42
& 6.42 & 2.68 & 0.45 \\
VidTIMIT
& 2.30 & 0.35 & 0.98 & 2.41
& 6.03 & 2.68 & 0.45 \\
\bottomrule
\end{tabular}%
}
\end{table}

\begin{table*}[!th]
\caption{Efficiency evaluation of secure matching under
different datasets, baselines, and network settings (RQ4).
Ratios are computed relative to \sysname{} within each
dataset. Best results are highlighted in bold.}
\centering
\begin{adjustbox}{max width=\textwidth}
\begin{tabular}{l|l|cc|cc|cc|cc|cc|cc|cc|cc}
\hline
\multirow{2}{*}{Dataset} & \multirow{2}{*}{Baseline}
& \multicolumn{4}{c|}{Comm. (MB)}
& \multicolumn{4}{c|}{Rounds}
& \multicolumn{4}{c|}{LAN Lat. (ms)}
& \multicolumn{4}{c}{WAN Lat. (ms)} \\
\cline{3-18}
& & Offline & Ratio & Online & Ratio
  & Offline & Ratio & Online & Ratio
  & Offline & Ratio & Online & Ratio
  & Offline & Ratio & Online & Ratio \\
\hline

\multirow{6}{*}{\begin{tabular}{c}
MULB+\\($K=7$,\\$d=3584$)
\end{tabular}}
& MP-SPDZ~\cite{keller2020mp}
& 1986.69 & 849.94 & 0.975 & 458.18
& 466 & 14.12 & 98 & 2.45
& 26512.54 & 15.27 & 57.29 & 3.57
& 48581.01 & 20.82 & 2124.30 & 2.60 \\

& PreprocessingforLife~\cite{boyle2025preprocessing}
& 1023.34 & 437.80 & 7.164 & 3366.54
& 88 & 2.67 & 217 & 5.42
& 47276.33 & 27.23 & 6074.72 & 378.18
& 57348.64 & 24.58 & 10503.82 & 12.85 \\

& MD-SONIC~\cite{zhang2025md}
& 4.011871 & 1.72 & 0.979184 & 460.14
& \textbf{13} & 0.39 & 44 & 1.10
& 4350.313 & 2.51 & 942.232 & 58.66
& 4559.300 & 1.95 & 1561.820 & 1.91 \\

& SHARK~\cite{gupta2025shark}
& 2.901070 & 1.24 & 0.462846 & 217.50
& 28 & 0.85 & 106 & 2.65
& 1661.665 & 0.96 & 29.459 & 1.83
& 2225.098 & 0.95 & 2162.000 & 2.65 \\

& LIGHTSHARK~\cite{zeng2026lightshark}
& \textbf{1.254470} & 0.54 & 0.004382 & 2.06
& 28 & 0.85 & 106 & 2.65
& \textbf{1659.265} & 0.96 & 27.175 & 1.69
& \textbf{2221.039} & 0.95 & 2152.450 & 2.63 \\

& \sysname{}
& 2.337437 & 1.00 & \textbf{0.002128} & 1.00
& 33 & 1.00 & \textbf{40} & 1.00
& 1736.107 & 1.00 & \textbf{16.063} & 1.00
& 2333.197 & 1.00 & \textbf{817.308} & 1.00 \\

\hline

\multirow{6}{*}{\begin{tabular}{c}
VidTIMIT\\($K=3$,\\$d=512$)
\end{tabular}}
& MP-SPDZ~\cite{keller2020mp}
& 468.74 & 1560.23 & 0.133 & 71.05
& 298 & 9.03 & 90 & 2.25
& 7556.72 & 5.05 & 36.63 & 2.31
& 17287.97 & 7.95 & 1927.33 & 2.36 \\

& PreprocessingforLife~\cite{boyle2025preprocessing}
& 111.72 & 371.87 & 0.820 & 438.03
& 85 & 2.58 & 216 & 5.40
& 5861.79 & 3.92 & 632.20 & 39.90
& 8681.96 & 3.99 & 5041.26 & 6.17 \\

& MD-SONIC~\cite{zhang2025md}
& 0.528115 & 1.76 & 0.108216 & 57.81
& \textbf{13} & 0.39 & 41 & 1.02
& 2542.380 & 1.70 & 476.819 & 30.09
& 2832.766 & 1.30 & 1201.555 & 1.47 \\

& SHARK~\cite{gupta2025shark}
& 0.378350 & 1.26 & 0.051678 & 27.61
& 28 & 0.85 & 90 & 2.25
& \textbf{1376.804} & 0.92 & 24.135 & 1.52
& 2050.925 & 0.94 & 1829.033 & 2.24 \\

& LIGHTSHARK~\cite{zeng2026lightshark}
& \textbf{0.193186} & 0.64 & 0.002686 & 1.43
& 28 & 0.85 & 90 & 2.25
& 1395.983 & 0.93 & 23.785 & 1.50
& \textbf{2034.355} & 0.93 & 1827.172 & 2.24 \\

& \sysname{}
& 0.300431 & 1.00 & \textbf{0.001872} & 1.00
& 33 & 1.00 & \textbf{40} & 1.00
& 1495.689 & 1.00 & \textbf{15.845} & 1.00
& 2175.864 & 1.00 & \textbf{816.850} & 1.00 \\

\hline
\end{tabular}
\end{adjustbox}
\label{tab:rq4_efficiency}
\end{table*}

This experiment tests that secure execution computes exactly the fixed-point decision on which policies are calibrated (Lemma~\ref{lem:cmp} and Section~\ref{ssec:design-range}). We compare, for every test trial, the decision in floating point on the unit vectors with the decision in plaintext on the fixed-point encodings. \sysname must reproduce the fixed-point decision on every trial. We choose $f=16$: with $d\le3584$ we have $\sqrt d\le60\le2^{f+1}$, so $\ell=64\ge2f+5$ places every compared value in range. Table~\ref{tab:rq3} shows that this precision changes almost no decisions. Among 99{,}834 MULB+ decisions, 99{,}833 agree with the fixed-point reference, with only one disagreement, while all 8{,}904 VidTIMIT decisions agree. In contrast, \sysname{} achieves exact agreement with the fixed-point reference on all 99{,}834 MULB+ and 8{,}904 VidTIMIT decisions, with zero disagreements. On the cosine scale, the maximum absolute score differences between floating-point and fixed-point execution are $2.8\times10^{-5}$ for MULB+ and $2.1\times10^{-5}$ for VidTIMIT. The minimum score-to-threshold distances for floating-point and fixed-point execution are $5.8\times10^{-6}$ and $1.9\times10^{-6}$ on MULB+, and $4.4\times10^{-6}$ and $1.2\times10^{-5}$ on VidTIMIT, respectively. For the full modality set, floating-point and fixed-point execution yield identical reported EER, FNMR, and FMR values: $0.47\%$, $0.31\%$, and $1.01\%$ on MULB+, and $0.00\%$, $0.00\%$, and $1.37\%$ on VidTIMIT. The identical decisions produced by \sysname{} also preserve these reported EER, FNMR, and FMR values.

\subsection{RQ4: Cost of Private Enforcement}
\label{ssec:eval-rq4}

Table~\ref{tab:rq4-plain} summarizes the plaintext authentication overhead in the single-sample setting. Feature extraction dominates the computation, accounting for nearly all of the 117.3~ms required for a MULB+ and 95.5~ms for a VidTIMIT. In VidTIMIT, the cost is primarily from processing eight face frames (86.7~ms), while in MULB+, hand extraction is the largest contributor (68.0~ms), followed by iris (29.1~ms) and face (10.0~ms). Fusion, encoding, and matching together contribute less than 0.1~ms per authentication attempt. Table~\ref{tab:rq3_part} presents the secure computation costs per authentication attempt. The server-side online latencies are 6.42~ms for MULB+ and 6.03~ms for VidTIMIT, with scoring (2.65~ms and 2.30~ms) and MAC checking (2.42~ms and 2.41~ms) accounting for most of the overhead. Selection takes 0.35~ms for both datasets, while comparison requires 0.99~ms and 0.98~ms, respectively. For both datasets, service-provider release and verification take 2.68~ms and 0.45~ms, respectively. Table~\ref{tab:rq4_efficiency} compares the communication
costs, rounds, and latency across different baselines under
LAN and WAN settings.
On MULB+ and VidTIMIT, \sysname{} incurs 2.34 and
0.30~MB of offline communication and 0.002 and
0.002~MB online, respectively,
requiring 33 offline and 40 online rounds on both datasets.
Its online latency is 16.62 and 16.16~ms under LAN,
and 819.58 and 817.82~ms under WAN.
Compared with SHARK, \sysname{}
achieves online speedups of 1.76$\times$ and 1.51$\times$
under LAN, and 2.65$\times$ and 2.24$\times$ under WAN.

%% file: sections/11-conclusion.tex
\section{Conclusion}
\label{sec:conclusion}
This paper studies modality flexibility in multi-biometric authentication from a security perspective. We show that allowing multiple authentication subsets expands the attack surface and that the resulting risk depends on their joint behavior rather than their individual performance.
To address this challenge, we develop a risk analysis framework and a policy selection method that balance security and usability requirements. We further present SUPERMBA, a maliciously secure two-server authentication system that privately enforces flexible authentication policies without revealing biometric templates or the modality subset used in an authentication attempt. 
Experiments on MULB+ and VidTIMIT demonstrate that modality flexibility can substantially increase authentication risk, while SUPERMBA achieves privacy-preserving policy enforcement with practical overhead. 
